\documentclass[article,nojss]{jss}

\usepackage{subcaption}
\usepackage{caption}

\usepackage{thumbpdf,lmodern}

\usepackage{framed}

\usepackage{amsmath,amsfonts}

\usepackage[normalem]{ulem}

\def\P{\mathbb{P}}
\newcommand{\di}{\text{\textnormal{d}}}
\newcommand{\e}{\text{\textnormal{e}}}
\newcommand{\be}{{\boldsymbol e}}
\newcommand{\bt}{{\boldsymbol\theta}}

\author{Sophie Hautphenne\\The University of Melbourne
   \And Brendan Patch\\The University of Melbourne}
\Plainauthor{Sophie Hautphenne, Brendan Patch}

\title{Birth-and-death Processes in \proglang{Python}:\\ The \pkg{BirDePy} Package}
\Plaintitle{Birth-and-death Processes in Python:\\ The BirDePy Package}
\Shorttitle{Birth-and-death Processes in \proglang{Python}: The \pkg{BirDePy} Package}

\Abstract{
 Birth-and-death processes (BDPs) form a class of continuous-time Markov chains that are particularly suited to describing the changes in the size of a population over time. 
 Population-size-dependent BDPs (PSDBDPs) allow the rate at which a population grows to depend on the current population size. 
 The main purpose of our new \proglang{Python} package \pkg{BirDePy} is to provide easy-to-use functions that allow the parameters of discretely-observed PSDBDPs to be estimated. 
 The package can also be used to estimate parameters of continuously-observed PSDBDPs, simulate sample paths, approximate transition probabilities, and generate forecasts. 
 We describe in detail several methods which have been incorporated into \pkg{BirDePy} to achieve each of these tasks. 
 The usage and effectiveness of the package is demonstrated through a variety of examples of PSDBDPs, as well as case studies involving annual population count data of two endangered bird species. 
}

\Keywords{ABC algorithm, Birth-and-death processes, continuous-time Markov chains, diffusion approximation, EM algorithm, Erlangization, Laplace transform, maximum likelihood estimation, parameter estimation, \proglang{Python}, uniformization}
\Plainkeywords{ABC algorithm, Birth-and-death processes, continuous-time Markov chains, diffusion approximation, EM algorithm, Erlangization, Laplace transform, maximum likelihood estimation, parameter estimation, Python, uniformization}

\Address{
  Sophie Hautphenne, Brendan Patch\\
  School of Mathematics and Statistics\\
  The University of Melbourne\\
  Victoria 3010, Australia\\
  E-mail: \email{\{sophiemh, patch\} @unimelb.edu.au}\\
  URL: \url{http://www.sophiehautphenne.info/}, \url{https://github.com/bpatch}
}

\begin{document}



\section[Introduction]{Introduction} \label{sec:intro}
 \emph{Birth-and-death processes} (BDPs) are continuous-time Markov chains that track the stochastically evolving number of individuals in a population over time \cite{feller1957introduction,grimmett2020probability}. 
These processes are effective modeling tools for areas such as biology, ecology, genetics, and epidemiology \citep{allen2008introduction,novozhilov2006biological,karlin2014first,naasell2002stochastic}, as well as  operations research \citep{brockmeyer1948erlang,boucherie2010queueing,gibbens1990bistability}. 
The state of a BDP transitions up by one unit when an individual is added to the population (e.g., due to a ``birth'' or ``arrival'') and transitions down by one unit when an individual is removed from the population (e.g., due to a ``death'' or ``departure''). 
\emph{Population-size-dependent BDPs} (PSDBDPs) allow realistic stochastic population dynamics to be modeled by letting the transition rates at any given time to depend upon the population size at that time (in a non-linear way). 
A recent review on PSDBDPs is provided by \cite{crawford2018computational}. 

This paper introduces a \proglang{Python} package, \pkg{BirDePy}, which is primarily intended to make a variety of parameter {\em estimation} tools for {\em discretely-observed} PSDBDPs available to biologists and other practitioners.  
\pkg{BirDePy} is the first software package to contain a comprehensive set of methods for performing this estimation task for PSDBDPs. 
Approaching the estimation problem from multiple angles is important since, in general, different circumstances call for different methods; for example, \textit{diffusion-approximation methods} may only be appropriate when population sizes are large, and \textit{matrix-exponential  methods} may only be possible when population sizes are small. 
In addition to targeting this estimation problem, \pkg{BirDePy} provides a complete set of tools for the computational analysis of PSDBDPs: it can also perform parameter estimation for \emph{continuously-observed} PSDBDPs, {\em simulate} discretely- and continuously-observed sample paths, compute approximations to {\em transition probabilities}, and generate {\em forecasts} of future expected behavior. 

Effective and accessible tools for estimating the parameters of PSDBDPs are crucial for the successful application of these processes. 
For continuously-observed processes in which the birth and death rates are linear functions of the unknown parameters,
closed form estimators are available \citep{wolff1965problems,reynolds1973estimating}. 
However, in practice observations are usually only available at discrete (often unequally-spaced) times. 
This makes the estimation task much more difficult since it is not at first sight clear how to account for the path that the process takes between observations.  
Furthermore, even if continuous observation is possible, despite a large and growing literature, determining exact estimators for realistic models that account for features such as logistic growth and the effect of a carrying capacity appears to be a highly non-trivial task. 
\cite{tavare2018linear} and \cite{asanjarani2021survey} provide recent surveys on parameter estimation for models related to PSDBDPs. 
We will soon return to parameter estimation for PSDBDPs after a short discussion on transition probabilities. 

Research focused on determining the probability of transition between any pair of states after any given time in PSDBDPs provides a foundation for parameter estimation, as these transition probabilities are used to compute likelihood functions. 
In often disparate work, researchers have approached the problem of determining transition probabilities using numerical methods. 
For example, it is well known that for models with a finite or truncated state space, the matrix exponential method can be used for this purpose \citep{feller1957introduction}. 
Other key advances in this area utilize Laplace transforms and continued fractions \citep{murphy1975some,crawford2012transition}, numerical approximations such as uniformization \citep{van2018uniformization,grassmann1977transient},  Erlangization \citep{asmussen2002erlangian,mandjes2016running}, and diffusion approximations \citep{kurtz1971limit,whitt2002stochastic}. 
Furthermore, research on determining transition probabilities has been explicitly connected to \textit{maximum-likelihood-estimates} (MLEs) for PSDBDPs \citep{ross2006parameter,ross2007estimation,ross2009parameter,buckingham2018gaussian}. 

The \textit{expectation-maximization} (EM) algorithm of \cite{dempster1977maximum} is an approach to obtaining MLEs in the presence of unobserved data. 
\cite{crawford2014estimation} apply the EM algorithm by
treating the (unobserved) evolution of a PSDBDP between discrete observation points as missing data. 
There are several ways to compute the expected value of the unobserved data, conditional on the observed data: \cite{crawford2014estimation} use a Laplace transform approach, but it can also be done using numerical integration, or by evaluating appropriate matrix exponentials. 
\pkg{DOBAD} is an \proglang{R} package related to this work written by \cite{DOBAD}. Recently, \cite{roney2020estipop} released the \proglang{R}-wrapped \proglang{C++} package \pkg{EstiPop} for simulation and parameter estimation based on diffusion-approximation, of a class of models that includes PSDBDPs. 
\pkg{StochSS} is a software-as-a-service platform with a graphical user interface that can be used to perform simulation of PSDBDPs \citep{drawert2016stochastic}; it can access the \proglang{Python} package \pkg{sciope} \citep{singh2021scalable} to use \textit{approximate Bayesian computation} (ABC) to estimate parameters for PSDBDPs. 
\cite{king2016statistical} developed \pkg{pomp}, which is an \proglang{R} package for working with partially observed Markov processes (POMPs); depending on the question and data at hand, POMPs and PSDBDPs can be used to address related research questions. 

As will be elaborated on later in the paper, \pkg{BirDePy} includes the EM algorithm, methods based on diffusion-approximation, and ABC parameter estimation methods akin to those in these other packages. 
It also incorporates all of the methods for finding transition probabilities for PSDBDPs mentioned earlier, it can use these to obtain MLEs, and it incorporates least-squares based estimation procedures. 
In addition to the Laplace transform based EM algorithm of \cite{crawford2014estimation}, \pkg{BirDePy} includes EM algorithm implementations that use numerical and matrix exponential based procedures for computing the required expected values. 
Additionally, \pkg{BirDePy} includes options to accelerate the EM algorithm \citep{jamshidian1997acceleration}. 
As well as being able to determine transition probabilities and perform parameter estimation, our package has several simulation algorithms built into it. 
The package can perform exact simulation \citep{kendall1950artificial,gillespie1977exact}, tau-leaping \citep{gillespie2001approximate,anderson2011error}, and a recently introduced piecewise approximation algorithm \citep{hautphenne2021modsim}. 
Another key strength of our package is that it includes a variety of built-in PSDBDP models, making it extremely easy to use. 
Due to this, all of the methods mentioned thus far can be called using a single line of code (after \pkg{BirDePy} has been imported). 
Finally, our package is well documented online \citep{birdepy}, further increasing its accessibility. 

The remainder of this paper is organized as follows. 
Section~\ref{sec:framework} formally introduces PSDBDPs along with the simulation, estimation, and forecasting methods used by \pkg{BirDePy}. 
The description and usage of \pkg{BirDePy} are given in Section~\ref{sec:software}. 
Section~\ref{sec:examples} demonstrates the capability of the package through a range of numerical examples, and Section~\ref{sec:case} applies the package to two topical examples involving endangered bird species. 
In the final section we discuss future possible enhancements to the package.

\section[Birth-and-death processes]{Birth-and-death processes}\label{sec:framework}
In this section we briefly summarize the theory underpinning the functionality of \pkg{BirDePy}. 
In Section~\ref{sec:models} we describe birth-and-death processes, in Section~\ref{sec:simulate} we provide simulation algorithms, in Section~\ref{sec:probability} we summarize methods for computing transition probabilities, and in Section~\ref{sec:estimate} we focus on parameter estimation frameworks.

\subsection{Models}\label{sec:models}
\pkg{BirDePy} is built for analyzing general birth-and-death processes. 
These continuous-time Markov chains (CTMCs) are widely used to study how the number of individuals in a population changes over time.
Let $Z(t)$ denote the size of a population at time $t\ge0$; the process $Z\equiv (Z(t),\,t\ge0)$ evolves on the state space $\mathcal S = (0,1,\dots,N)$, where $N$ may be infinite. 
We assume that the evolution of such a process is fully described by non-negative real-valued functions $\lambda_z(\bt) : \mathcal S \to \mathbb R_+$  and $\mu_z(\bt) : \mathcal S \to \mathbb R_+$ of the current state $z$; these functions are known as the \emph{population ``birth-rate''} and  \emph{``death-rate''}, respectively. 
The model has a finite number of real parameters recorded in the vector $\bt$. 
When $Z(t)=z$, the time until the process transitions to another state is exponentially distributed with mean $(\lambda_z(\bt)+\mu_z(\bt))^{-1}$ (where we define $1/0=\infty$). 
At a transition time from state $z$, the process jumps to state $z+1$ with probability $\lambda_z(\bt)/(\lambda_z(\bt)+\mu_z(\bt))$, or to state $z-1$ with probability $\mu_z(\bt)/(\lambda_z(\bt)+\mu_z(\bt))$. 

As summarized in Table~\ref{tab:rates}, we consider selected birth-and-death models of three types: (i) some variants of the linear BDP, (ii) some PSDBDPs, and (iii) some queueing models.  
We give a brief description of these models in the next paragraphs.

\begin{table}[h!]
  \begin{center}
    \begin{tabular}{p{3.5cm}|c|c|p{2.2cm}} 
      \textbf{Model label} & $\lambda_z(\bt)$ & $\mu_z(\bt)$  & $\bt$\\
      \hline\hline
      \code{"linear"} & $\gamma z$ & $\nu z$ & $\gamma$, $\nu$\\
      \code{"linear-migration"} & $\gamma z + \alpha$ & $\nu z$ & $\gamma$, $\nu$, $\alpha$ \\
      \code{"pure-birth"} & $\gamma z$ & 0 & $\gamma$\\
      \code{"pure-death"} & 0 & $\nu z$ & $\nu$\\
      \code{"Poisson"} & $\gamma$ & 0 & $\gamma$\\
      \hline
      \code{"Verhulst"}& $\gamma\left(1-\alpha z\right) z$ & $\nu\left(1+\beta z\right)z$ & $\gamma$, $\nu$, $\alpha$, $\beta$\\
      \code{"Ricker"} & $\gamma z \exp\left(-(\alpha z)^c \right)$ & $\nu z$ & $\gamma$, $\nu$, $\alpha$, $c$\\
      \code{"Hassell"} & $\frac{\gamma z}{(1+\alpha z)^c}$ & $\nu z$ & $\gamma$, $\nu$, $\alpha$, $c$\\
      \code{"MS-S"} & $\frac{\gamma z}{1+ \left(\alpha z\right)^c}$ & $\nu z$ & $\gamma$, $\nu$, $\alpha$, $c$ \\
      \code{"Moran"} & $\frac{N-z}{N}\left(\frac{\alpha z(1-u) + \beta (N-z)v}{N}\right)$ & $\frac{z}{N}\left(\frac{\beta(N-z)(1-v) + \alpha z u}{N}\right)$ & $\alpha$, $\beta$, $u$, $v$, $N$\\
      \hline
      \code{"M/M/1"} & $\gamma$ & $\nu1_{\{z>0\}}$ & $\gamma$, $\nu$\\
      \code{"M/M/inf"} & $\gamma$ & $\nu z$ & $\gamma$, $\nu$\\    
      \code{"loss-system"} & $\gamma1_{\{z < c\}}$ & $\nu z$ & $\gamma$, $\nu$, $c$
    \end{tabular}
        \caption{Transition rates and parameters for $z \geq 0$. Any rates which take on a negative value are modified to be equal to 0, and $1_{\{A\}}$ is the indicator of the event $A$. The order of the parameters in the last column is important for Section~\ref{sec:software}. }
    \label{tab:rates}
  \end{center}
\end{table}

A \emph{linear} BDP models the evolution of a population in which individuals give birth and die independently of each other and of the current population size.  
The birth and death rates \emph{per individual} are given by some strictly positive constants $\gamma$ and $\nu$, respectively, leading to linear \textit{population} birth and death rates $\lambda_z(\boldsymbol{\theta})=\gamma z$ and $\mu_z(\boldsymbol{\theta})=\nu z$. 
If, in addition, individuals immigrate at rate $\alpha$ into the population, the birth rate becomes $\lambda_z(\boldsymbol{\theta})=\gamma z+\alpha$.  
A pure birth process, also called \emph{Yule} process, is obtained by setting the death rate $\nu$ to zero, and a pure death process is obtained by setting the birth rate $\gamma$ to zero.
The \emph{Poisson process} is a particular pure birth process in which the birth rate $\lambda_z(\boldsymbol{\theta})$, also called \emph{arrival} rate, does not depend on the current population size $z$. Classical references discussing these models include \cite{cinlar2013introduction} and  \cite{karlin2014first}.

In contrast to linear birth-and death processes, in a PSDBDP the birth and death rates per individual depend on the current population size $z$. 
This feature allows us to model competition for resources such as food or territory. The birth (respectively death) rate per individual is typically a decreasing (respectively increasing) function of $z$; this leads to \emph{logistic growth} of the population (S-shaped trajectories), which tend to fluctuate around a threshold value called the \emph{carrying capacity}. 
One of the most popular logistic models is the \emph{Verhulst model}, which has been reinvented several times since it first appeared in 1838 (see for instance \cite{naasell2001extinction} and references therein). 
In this model, the birth and death rates per individual are linear functions of $z$. 
The \emph{Susceptible-Infectious-Susceptible} (SIS) \emph{epidemic model} is a particular case of the Verhulst model with $\beta=0$, implying the death (recovery) rate per individual does not depend on $z$.  
Other classical models of logistic growth include the \emph{Ricker model}, in which the birth rate decreases exponentially in $z$, and the \emph{Hassell} and \emph{MaynardSmith-Slatkin} (MS-S) models, in which the birth rate decreases according to a power law (here we assume that the death rate in these models is  independent of the population size).  
Despite the apparent similarity between the Hassell and the MS-S models, the MS-S model has a more flexible and better descriptive form than the Hassell model according to \cite{bellows1981descriptive}. 
An overview of several of these logistic models in the deterministic case is given by \cite{jemmer2005discrete}.
The particular case where $c=1$, called the \emph{Beverton-Holt} (B-H) model, is also widely used, in particular as a model for exploited fish populations \citep{subbey2014modelling}. Finally, the \emph{Moran process} models the change in the numbers of particles of two types in a finite population of size $N$, where transitions between types occur at a rate proportional to the number of potential contacts between members of each type in a population; this is an important model for genetic evolution, see for example \cite[Chapter 6]{nowak2006evolutionary}.

In classical queueing models (see \cite{kleinrock1975queuing,karlin1981second}), the population birth rate $\lambda_z(\boldsymbol{\theta})$ does not depend on $z$: births (arrivals) occur according to a point process, and once in the population (system), individuals queue to be served by one or multiple servers. 
The \emph{M/M/1} and \emph{M/M/$\infty$ queues} have arrivals occurring according to a Poisson process and, respectively, a single server and an infinite number of servers (acting in parallel and independently). 
In a \emph{loss system}, arrivals can only occur when the population is below a certain threshold value $c$.

In the remainder of this paper, when the dependence on $\bt$ is clear from the context, we simply write $\lambda_z(\bt)$ and $\mu_z(\bt)$ as $\lambda_z$ and $\mu_z$. 

\subsection{Simulation}\label{sec:simulate}
A few exceptions aside, explicit expressions for quantities of interest such as the probability distribution, or the expected value, of the population size of a PSDBDP at a given time, do not exist. By providing many examples of possible outcomes,
Monte Carlo simulation can allow interesting questions related to these processes to be examined (e.g., \cite{karev2003simple,karev2004gene}). 
Simulation can also be used to compare the performance of other tools, such as those used for parameter estimation, by providing synthetic data for the tools to be tested on (e.g., \cite{davison2021parameter,mandjes2021numerical}). 
Last, but not least, some parameter estimation techniques directly rely on simulated sample paths to generate estimates (e.g., \cite{singh2021scalable}). 

Simulation of a birth-and-death process can be {continuous} or {discrete}. The goal of \emph{continuous} simulation is
 to obtain a sequence of times at which the process transitions, together with the corresponding sequence of states the process transitions into. 
In a continuous simulation, the output provides sufficient information to recover the state of the process at any time within the simulation horizon. 
This is desirable since the output contains enough information for a user to know whether certain events have occurred, for example whether the process has crossed a certain boundary by a certain time. 
In contrast, the goal of {\em discrete} simulation is
 to obtain samples of a birth-and-death process at pre-specified observation times. 
In this case, the state of the process at those observation times is the output.
An advantage of discrete simulation is that the same observation times can be specified for each sample path so that they can be combined together to find, for example, the expected value or variance of a process as a function of time. 

Given an initial population size $z_0$, to perform continuous simulation of a birth-and-death process over the interval $[0,t]$, initialize $j:=0$, $t_0:=0$, and then repeat these steps:
\begin{enumerate}
\item Generate an outcome $\Delta$ of an exponentially distributed random variable with mean $(\lambda_{z_j}+\mu_{z_j})^{-1}$.
\item If $t_j+\Delta\le t$, set $t_{j+1}:=t_j+\Delta$; otherwise stop. 
\item With probability $\lambda_{z_j}/(\lambda_{z_j}+\mu_{z_j})$ set $z_{j+1}:=z_j+1$; otherwise set $z_{j+1}:=z_j-1$.
\item Set $j:=j+1$. 
\end{enumerate}
Discrete simulation proceeds in a similar fashion. 
Given an initial population size $z_0$, to simulate a birth-and-death process at the time points $t_0<t_1<\dots <t_n$, initialize $j:=1$, $s:=t_0+\Delta$ where $\Delta$ is an outcome of an exponentially distributed random variable with mean $(\lambda_{z}+\mu_{z})^{-1}$, and $z:=z_0$, and then repeat these steps:
\begin{enumerate}
\item While $s \le t_j$:
  \begin{enumerate}
    \item[(i)] With probability $\lambda_{z}/(\lambda_{z}+\mu_{z})$ set $z:=z+1$; otherwise set $z:=z-1$.
    \item[(ii)] Set $s := s +\Delta$ where $\Delta$ is an outcome of an exponentially distributed random variable with mean $(\lambda_{z}+\mu_{z})^{-1}$.
  \end{enumerate} 
\item Set $z_{j}:=z$. 
\item If $j<n$ set $j:=j+1$; otherwise stop.
\end{enumerate}
Observe that $t_j$ has a different interpretation depending on whether continuous or discrete simulation is being performed.  
It is thought within the probability community that this {\em exact} approach to simulation of CTMCs was first considered by Joseph L.\ Doob and his collaborators around 1945. 
To our knowledge, the first implementation on a computer is mentioned in \cite{kendall1950artificial}, and the algorithm was then popularized more widely in \cite{gillespie1977exact}. 

A well-known drawback of exact simulation algorithms is that they may take considerable computational time to produce sample paths. 
An alternative \emph{approximate} simulation method for population processes known as {\em tau-leaping}, introduced by \cite{gillespie2001approximate}, allows the user to trade accuracy for speed. 
This algorithm partitions time into intervals of predetermined constant length $\tau$.  
Conditional on $Z(0)=z$ at the start of an interval, the state of $Z(\tau)$ is approximated by the difference of two Poisson distributed random variables $\mathfrak L$ and $\mathfrak M$ which are respectively intended to approximate the total number of births and deaths that occur within the interval. 
Given an initial population size $z_0$, the basic {\em Euler} form of the algorithm can be used to perform discrete simulation by initializing $j:=1$, $s:=t_0+\tau$, and $z:=z_0$, and then repeating these steps:
\begin{enumerate}
  \item While $s\le t_j$:
    \begin{enumerate}
    \item[(i)] Set $z := z +\mathfrak L -\mathfrak M$ where $\mathfrak L$ and $\mathfrak M$ are are outcomes of Poisson distributed random variables with respective means $\lambda_{z}\tau$ and $\mu_{z}\tau$. 
    \item[(ii)] Set $s:=s+\tau$. 
    \end{enumerate}
\item Set $z_{j}:=z$. 
\item If $j<n$ set $j:=j+1$; otherwise stop.  
\end{enumerate}
Several variations of this algorithm have been developed which focus primarily on step-size selection (e.g., \cite{gillespie2003improved,cao2006efficient}) and on ensuring that the population size does not go negative during the simulation (e.g., \cite{cao2005avoiding,chatterjee2005binomial,anderson2008incorporating}). 

In \cite{anderson2011error} a {\em midpoint} variant of the basic algorithm is developed and elegantly analyzed. 
The  algorithm is the same as the Euler version described above with the exception that the two Poisson distributed random variables $\mathfrak L$ and $\mathfrak M$ have respective means $\lambda_{z+\rho(z)}\tau$ and $\mu_{z+\rho(z)}\tau$, where
 $\rho(z)=\frac{1}{2}\tau\big(\lambda_z-\mu_z\big)$ approximates $\mathbb E [Z(\frac{1}{2}\tau)-Z(0)\,|\,Z(0)=z]$. 
The basic idea here is that determining the number of births and deaths using an approximation to the population size at the midpoint of an interval may increase accuracy relative to using the population size at the beginning of the interval. 
This algorithm can provide substantial improvements in accuracy relative to the Euler version at only a minor increase in computational cost. 
Both of the above algorithms can be thought of as generating piecewise approximations to a PSDBDP, where zero-order approximations of the birth and death rates are utilized within each subinterval.

Recently a new approach to perform discrete simulation of a general birth-and-death processes utilizing linear (first-order) approximations of the birth and death rates within each subinterval was proposed in \citep{hautphenne2021modsim}. 
This algorithm approximates any birth-and-death process by `piecewise-linear' birth-and-death processes. 
It also uses the fact that 
a discretely-observed linear birth-and-death process corresponds to a {\em linear fractional Galton--Watson process} (see \cite{harris1963theory}). 
This means that for a linear birth-and-death process $\mathring Z$ with birth rates $\lambda_z=\lambda z$ and death rates $\mu_z=\mu z$, the probability that a family generated by a single individual at time $0$
 `becomes extinct' before time $\tau$ is given by 
\begin{equation}\label{eq:sim_survive}
\beta_1(\tau) = \left\{\begin{array}{cc} \mu\{\exp\big((\lambda-\mu)\tau\big)-1\}/\{\lambda \exp\big((\lambda-\mu)\tau\big)-\mu\} & \text{if } \lambda\ne \mu, \\ 
\lambda \tau/(1+\lambda \tau) & \text{if } \lambda = \mu. 
\end{array}\right.
\end{equation}
Each family which survives results in $H$ individuals being present at time $\tau$, where $\mathbb P(H=h) = (1-\beta_2(\tau))\beta_2(\tau)^{h-1}$ with 
\begin{equation}\label{eq:binom_success}
\beta_2(\tau) = \left\{ \begin{array}{cc} \lambda\beta_1(\tau)/\mu& \text{if } \lambda\ne \mu, \\ 
\beta_1(\tau) & \text{if } \lambda = \mu. 
\end{array}\right.
\end{equation} 
Therefore, given $\mathring Z(0)=z$, a realization of $\mathring Z(\tau)$ can be obtained by generating an outcome of the random variable  $\mathfrak B$  
which is binomially distributed with $z$ trials and success probability $1-\beta_1(\tau)$ given by \eqref{eq:sim_survive}, and then using the fact that $\mathring Z(\tau)$ follows a negative binomial distribution with parameters $\mathfrak B$ and $1-\beta_2(\tau)$, that is,
\begin{equation}\label{eq:neg_binom}
\mathbb P(\mathring Z(t_j) = k\,|\,\mathfrak B) = \binom{k+\mathfrak B-1}{k}\beta_2(\tau)^{\mathfrak B}(1-\beta_2(\tau))^k. 
\end{equation}

So, if $Z$ is a PSDBDP, conditional on $Z(0)$, $Z(\tau)$ can be approximated by $\mathring Z(\tau)$
by setting the per-individual rates of the approximating linear process $\mathring Z$ equal to those of $Z$ at time $0$, that is, $\lambda = \lambda_{Z(0)}/Z(0)$ and $\mu=\mu_{Z(0)}/Z(0)$. 
Therefore, to obtain an approximate simulation of $(Z(t_j),~j=0,1,\dots,n)$, initialize $j:=1$, $s:=t_0+\tau$, and $z:=z_0$ , and repeat these steps:
\begin{enumerate}
\item While $s\le t_j$:
  \begin{enumerate}
    \item[(i)] Set $\lambda := \lambda_{z}/z$ and $\mu := \mu_{z}/z$. 
    \item[(ii)] Generate a binomially distributed random variable $\mathfrak B$ with success probability $1-\beta_1(\tau)$ with $\beta_1(\tau)$ as given by \eqref{eq:sim_survive} and number of trials $z$.    
    \item[(iii)] Generate a negatively binomially distributed random variable $\mathfrak C$ with success probability $1-\beta_2(\tau)$ with $\beta_2(\tau)$ as given by \eqref{eq:binom_success} and number of trials $\mathfrak B$. 
    \item[(iv)] Set $z := z +\mathfrak C$ and $s:=s+\tau$.
  \end{enumerate}
\item Set $z_j:=z$.
\item If $j<n$ set $j:=j+1$; otherwise stop.  
\end{enumerate}
This algorithm can be implemented efficiently and is highly accurate. 
It also explicitly avoids the possibility of the population size becoming negative during the simulation. 

Table~\ref{tab:sims} summarizes the discrete simulation methods described in this section, and gives the label used to call them in the \pkg{BirDePy} function \code{birdepy.simulate.discrete()} as described in Section~\ref{sec:software}. 

\begin{table}[h!]
  \begin{center}
    \begin{tabular}{p{5.4cm}|c|p{7cm}} 
      \parbox[t]{5cm}{\textbf{Method}\\ \textbf{(in order of appearance)}} & \textbf{Label} & \textbf{Brief description}\\
      \hline
      Exact& \code{"exact"} & Utilizes all jumps of the process.\\
     Euler approximation & \code{"ea"} & Population changes between $\tau$-sized intervals governed by Poisson random variables with parameters depending on population sizes at beginning of intervals.\\
      Midpoint approximation & \code{"ma"} & Population changes between $\tau$-sized intervals governed by Poisson random variables with parameters depending on estimate of population sizes at midpoints of intervals.\\
     Galton--Watson approximation & \code{"gwa"} & Population changes between $\tau$-sized intervals governed by linear birth-and-death processes with parameters depending on population sizes at beginning of intervals.\\
    \end{tabular}
        \caption{Methods for simulating sample paths of general birth-and-death processes.}
    \label{tab:sims}
  \end{center}
\end{table}

\subsection{Transition probabilities}\label{sec:probability}
Many of the general models described in Section~\ref{sec:models} have
no explicit expression for their transition probabilities
$p_{i,j}(t) := \P(Z(t)=j\mid Z(0)=i),$ $i,j\in \mathcal S$, $t\geq 0$. 
However, it is often desirable to compute these transition probabilities since they allow for a deeper understanding of the future (random) evolution of the process. 
Practically speaking, this may underpin some performance analysis of a system which is being modelled by the process. 
Additionally, and of most interest in this paper, transition probabilities allow likelihood functions of discretely-observed sample paths to be evaluated, which opens up the possibility of parameter estimation (as discussed in the next section). 
In this section we outline nine methods of approximating $p_{i,j}(t)$. 
Three of these methods are matrix-based and rely on state space truncation, another method uses Laplace transforms, and the remaining four methods use approximation models. 

\subsubsection{Matrix exponential}
The Kolmogorov forward equation (e.g., \cite[Section~6.9]{grimmett2020probability}) for CTMCs provides the foundational property that the collection of transition probability functions $p_{i,j}(t)$ satisfy, for $t\ge0$, the system of first order ordinary differential equations
\begin{equation}\label{eq:kol}
\begin{split}
\frac{ \di p_{i,0}(t)}{\di t} &= \mu_1 p_{i,1}(t) - \lambda_0p_{i,0}(t),\\
\frac{ \di p_{i,j}(t)}{\di t} &= \lambda_{j-1}p_{i,j-1}(t) + \mu_{j+1}p_{i,j+1}(t)- (\lambda_j+\mu_j)p_{i,j}(t), \quad j\ge 1,\\
\end{split}
\end{equation}
with $p_{i,i}(0) = 1$ and $p_{i,j}(0)=0$ for $j\ne i$. 
Let $Q$ be the \emph{generator }of $Z$, that is, the square matrix with diagonal entries $q_{i,i}=-(\lambda_i+\mu_i)$, upper diagonal entries $q_{i,i+1}=\lambda_i$, and lower diagonal entries $q_{i-1,i}=\mu_i$ (and zeros elsewhere). 
Also collect the transition probabilities into a matrix $P(t) = \big(p_{i,j}(t); i,j\in\mathcal S\big)$. 
Then \eqref{eq:kol} can be written in matrix form as $\frac{\di}{\di t} P(t) = P(t)Q$. 
When $Q$ is finite it immediately follows that 
\begin{equation}\label{eq:matrix_exp}
P(t) = \lim_{k\to\infty}\sum_{n=0}^k \frac{1}{n!}(Qt)^{n} =: \exp(Qt).
\end{equation} 

Matrix exponentials are ubiquitous and their efficient computation has received a great deal of attention over the years; see \cite{moler2003nineteen} for a review, and \cite{al2010new} for a method in popular usage. 
In particular, efficient methods for determining an appropriate truncation of the infinite series in \eqref{eq:matrix_exp} and approximating the remaining tail sum are available. 
Despite this attention, there have been doubts about the effectiveness of this approach when $Q$ becomes large or nearly singular \cite{ross2006parameter,crawford2012transition}. 
In addition, to use this method to compute transition probabilities, the maximum possible population size must be bounded ($|\mathcal S|<\infty$). 
For models where this does not naturally occur, $Q$ can be truncated to obtain approximate results. 
A modern perspective on the utility of the matrix exponential for computing transition probabilities of finite-state CTMCs can be found in \citep{sherlock2021direct}.

\subsubsection{Uniformization}
Due to the special nature of the generator $Q$ (specifically that it has non-negative off-diagonal entries and row-sums equal to 0), an alternative procedure known as {\em uniformization}, introduced by \cite{jensen1953markoff}, is available for computing $P(t)$.  
Consider a discrete-time Markov chain (DTMC) with probability transition matrix $A:=Q/a+I$, where $a:=\max_z |\lambda_z+\mu_z|$ when this exists (for example when $|\mathcal S|<\infty$), and where $I$ denotes the identity matrix of appropriate size.
Suppose that this DTMC transitions at the event times of a Poisson process $(N(t),~t\ge0)$ with rate $a$. 
Using this construction,  we can show that 
$$\P(Z(t)=j\mid Z(0)=i, N(t)=n)=(A^n)_{i,j}.$$
Therefore, conditioning on $N(t)$ provides
\begin{equation}\label{eq:uniform}
P(t) = \lim_{k\to\infty}\sum_{n=0}^k \frac{(ta)^n\e^{-ta}}{n!} A^n. 
\end{equation}
This procedure is discussed in detail in \cite{grassmann1977transient,gross1984randomization,melamed1984randomization}, and a recent review of the method is provided in \cite{van2018uniformization}.

In addition to potentially requiring truncation of the state space for many models of interest, another source of approximation error for this method is that a finite $k$ must be chosen in~\eqref{eq:uniform}. 

\subsubsection{Erlangization}
Using probabilistic arguments, {\em Erlangization} is yet another matrix-based method for computing $P(t)$. 
Let $T$ be an exponentially distributed random variable with mean $\eta^{-1}>0$. 
Define $r^{(\eta)}_{i,j} := \P(Z(T) = j\mid Z(0)= i)$ and collect these quantities into  $R(\eta) = \big(r^{(\eta)}_{i,j},~i,j\in \mathcal S)$. 
The matrix $R(\eta)$ can be viewed as the one-step transition probability matrix of a DTMC embedded in $Z$ at the epochs of a Poisson process with rate $\eta$. 
By conditioning on the first transition of $Z$ and the expiry of the time $T$, we obtain the recursive expressions 
\begin{equation}\label{eq:erlang}
r^{(\eta)}_{i,j}= \frac{\mu_ir^{(\eta)}_{i-1,j} + \lambda_ir^{(\eta)}_{i+1,j} + \eta 1_{\{i=j\}}}{\lambda_i + \mu_i + \eta},\quad i,j\in\mathcal S. 
\end{equation}
This system of equations can be written in matrix form in terms of $R(\eta)$, whose solution is given in terms of the generator $Q$ by
\[
R(\eta) = \eta\big(\eta I-Q)\big)^{-1}.
\]
Therefore, if we let $\eta:=k/t$ and $S_{k,t}$ be an Erlang distributed random variable with rate parameter $k/t$ and shape parameter $k$, then the $(i,j)$th entry of the matrix $R(k/t)^k$ contains $\mathbb P(Z(S_{k,t})=j\mid Z(0)=i)$.
The expected value of $S_{k,t}$ is $t$ and the variance of $S_{k,t}$ is $t/k$.
This means that
\begin{equation}\label{eq:erlangization}
P(t) = \lim_{k\to\infty}R(k/t)^k. 
\end{equation}

The Erlangization method for approximating transition probabilities is discussed in \cite{asmussen2002erlangian,mandjes2016running,stanford2011erlangian,mandjes2021numerical} for models related to birth-and-death processes. 
Similar to the uniformization method, error arises in the Erlangization method since the state space may need to be truncated,
and the infinite limit in Equation~\eqref{eq:erlangization} needs to be approximated by a finite $k$. 

\subsubsection{Inverse Laplace transform}
The Laplace transform of the transition function $p_{i,j}(t)$ is 
\begin{equation}\label{eq:laplace}
f_{i,j}(s) = \mathcal L[p_{i,j}](s) = \int_0^\infty p_{i,j}(t)\e^{-st} \di t. 
\end{equation} 
Let $\frac{u_1}{v_1+}\frac{u_2}{v_2+}\frac{u_3}{v_3+}\cdots$ be a short-hand notation for the \emph{continued fraction}
\[
\dfrac{u_1}{v_1+\dfrac{u_2}{v_2+\dfrac{u_3}{v_3 +\cdots }}},
\] 
where $(u_i,~i=1,2,\dots)$ and $(v_i,~i=1,2,\dots)$ are sequences of real numbers. 
As first reported in \cite{murphy1975some} and detailed in \cite{crawford2012transition}, the Laplace transform \eqref{eq:laplace} takes the continued fraction form
\begin{equation}\label{eq:laplace_ij}
f_{i,j}(s) = \left\{ \begin{array}{ll}
\left(\prod_{k=j+1}^i\mu_k\right)\frac{B_j(s)}{B_{i+1}(s)+}\frac{B_i(s)\,a_{i+2}}{b_{i+2}(s)+}\frac{a_{i+3}}{b_{i+3}(s)+}\cdots, & \text{for } j \le i,\\[0.5em]
\left(\prod_{k=i}^{j-1}\lambda_k\right) \frac{B_i(s)}{B_{j+1}(s)+}\frac{B_j(s)\,a_{j+1}}{b_{j+2}(s)+}\frac{a_{j+3}}{b_{j+3}(s)+}\cdots, & \text{for } j \ge i,
\end{array}\right.
\end{equation}
where 
\begin{align*}
B_0(s)&=1,\\
B_1(s)&=b_1(s),\quad\text{and}\\
B_k(s)&=b_k(s)B_{k-1}(s)+a_kB_{k-2}(s),\quad\text{for } k\ge 2,
\end{align*}
with $a_1=1$ and $a_j = -\lambda_{j-2}{\mu_{j-1}}$ for $j\ge2$, and $b_1(s)=s+\lambda_0$ and $b_j(s)=s+\lambda_{j-1}+\mu_{j-1}$ for $j\ge2$. 
An advantage of the continued fraction form of Equation~\eqref{eq:laplace_ij} is that it can be evaluated using the Lentz algorithm to a user-specified error tolerance \cite[Chapter~5]{press2007numerical}. 
The Laplace transform can then be numerically inverted to obtain $p_{i,j}(t)$. 

Numerical Laplace transform inversion has been the subject of intense research efforts over the past few decades. 
Some recent reviews include \cite{kuhlman2013review,wang2015different}, while \cite{davies1979numerical,abate2000introduction} provide earlier (yet thorough) discussion on the topic. 
Methods which have stood the test of time are presented in \cite{de1982improved,talbot1979accurate,stehfest1970algorithm,valko2004comparison}, a unified framework is developed in \cite{abate1995numerical,abate2006unified}, and a promising new approach is developed in \cite{horvath2020numerical}.  

\subsubsection{Diffusion approximation}
Define $(Z_i^{(r)},~r \in\mathbb N)$ to be a parametric family of CTMCs which evolve according to transition rates $\lambda_z^{(r)} = r\lambda_{z/r}$ and $\mu_z^{(r)} = r\mu_{z/r}$ with $Z_i^{(r)}(0)=i$. 
For $s \in[0, t]$ define the diffusion-scaled process 
\[
\tilde Z_i(s) = \lim_{r\to\infty}\sqrt{r}\left(\hat Z_i^{(r)}(s) - \hat z_i(s)\right),
\]
where $\hat Z_i^{(r)}:=\frac{1}{r} Z_i^{(r)}(s)$ and $\hat z_i(s)$ satisfies 
\[
\frac{ \di }{\di s}\hat z_i(s) = \lambda_{\hat z_i(s)} - \mu_{\hat z_i(s)},\qquad \hat z_i(0) = i.
\]
Loosely speaking, Theorem~3.5 in \cite{kurtz1971limit} can be used to show that, under some regularity conditions, $(\hat Z_i^{(r)}(s),~s\in[0,t])$ converges weakly, as $r\to\infty$, in the space of cadlag functions on $[0, t]$ to a zero-mean Gaussian diffusion $(\tilde Z_i(s),~s\in[0,t])$ with variance 
\[
\sigma_i^2(s) := \text{Var}(\tilde Z(s)) = M(s)^2\int_0^s \left(\lambda_{\hat z(\tau)} + \mu_{\hat z(\tau)}\right) M(\tau)^{-2} \di \tau
\] 
for each $s\in[0, t]$, 
where $M(s):= \exp\left(\int_0^s B(\tau)d \tau\right)$ with $B(\tau) = H\left(\hat z(\tau)\right)$ defined in terms of 
\[
H(z) = \frac{\di}{\di z}\Big(\lambda_z - \mu_z\Big).
\] 
In particular this implies that 
\[
p_{i,j}(t) \approx \mathbb P(\tilde Z_i(t)=j). 
\]
Hence $p_{i,j}(t)$ can be approximated by the probability density of a normally distributed random variable with mean $\hat z_i(t)$ and variance $\sigma_i^2(t)$ as given above.
 
This approach is very closely related to the functional central limit theorem (e.g., Theorem~4.3.2 in \cite{whitt2002stochastic}). 
The diffusion approach to approximate transition probabilities is used in \cite{ross2009parameter}, and discussed at length in \cite{allen2008introduction}. 

\subsubsection{Ornstein--Uhlenbeck approximation}
The diffusion approximation discussed above can be substantially simplified if it is assumed that $Z$ is fluctuating about a steady state point $z_{\text{eq}}$ of $\hat z$. 
Such a point occurs when the birth rate is equal to the death rate, i.e., $z_{\text{eq}}$ satisfies $\lambda_{z_{\text{eq}}} = \mu_{z_{\text{eq}}}$.
In this case, as argued in \cite{ross2006parameter}, the limiting Gaussian diffusion is an Ornstein--Uhlenbeck process. 
Therefore, $p_{i,j}(t)$ can be approximated by the density of a normally distributed random variable with mean $\tilde z= z_{\text{eq}} + e^{H(z_{\text{eq}})t}(i - z_{\text{eq}})$ and variance $\tilde\sigma^2 = \frac{\lambda_{z_{\text{eq}}} + \mu_{z_{\text{eq}}}}{2H(z_{\text{eq}})}\left(e^{2H(z_{\text{eq}})t}-1\right)$. 
This method assumes that $z_{\text{eq}}$ is asymptotically stable (i.e., $H(z_{\text{eq}})<0$), and as such favors values of $z_{\text{eq}}$ that minimize $H(z_{\text{eq}})$. 

\subsubsection{Galton--Watson approximation}
Recall from Section \ref{sec:simulate} that a PSDBDP can be approximated by a piecewise-linear birth-and-death process by decomposing time into sub-intervals of finite length, and letting the per-individual birth and death rates be constant over each time interval; more precisely, if $z$ is the population size at the beginning of a time interval, then we let $\lambda := \frac{1}{z}\lambda_{z}$ and $\mu := \frac{1}{z}\mu_{z}$ over that interval (in Section  \ref{sec:simulate} we assumed the intervals to be of constant length $\tau$, but this assumption is not necessary). We can extend this idea to obtain approximations for the transition probabilities in a PSDBP.

Suppose we want to approximate $p_{i,j}(t)$ for some $i,j\in\mathcal{S},$ and $t> 0$. Consider a linear birth-and-death process $\mathring Z^{(b)}$ with per-individual birth rate $\lambda = \frac{1}{b}\lambda_{b}$ and per-individual death rate $\mu = \frac{1}{b}\mu_{b}$, where possible choices of $b$ include $b=i$ (such as in Section  \ref{sec:simulate}), $b=j$, $b=\max(i,j)$, $b=\min(i,j)$, and $b=(i+j)/2$. 
The transition probability $p_{i,j}(t)$ can then be approximated by 
\[
p_{i,j}(t) \approx \mathring p_{i,j}(t) :=\mathbb P(\mathring Z^{(b)}(t)=j\mid \mathring Z^{(b)}(0)=i).
\]
What constitutes a good choice of $b$ will depend highly on $i$, $j$ and the model under study (this is a topic for future research). 
This approximation is particularly convenient since it is well known (see \cite{guttorp1991statistical}) that $\mathring p_{i,0}(t) = \beta_1(t)^{i}$, and for $j\ge1$,
\begin{equation}\label{eq:gw}
\mathring p_{i,j}(t) = \sum_{k=\max(0, i-j)}^{i-1} \binom{i}{k} \binom{j -1}{i-k-1} \beta_{1}(t)^k\Big[\big\{1-\beta_{1}(t)\big\}\big\{1-\beta_{2}(t)\big\}\Big]^{j-k} \beta_{2}(t)^{j-i+k},
\end{equation}
where $\beta_1(t)$ and $\beta_2(t)$ are given, respectively, by \eqref{eq:sim_survive} and \eqref{eq:binom_success}. 
When the binomial coefficients in \eqref{eq:gw} cause numerical problems or take a long time to compute, an alternative expression developed by \cite{davison2021parameter} using a saddlepoint approximation \cite{butler2007saddlepoint} may be used. 

\subsubsection{Monte Carlo simulation}
By using the simulation methods described in Section~\ref{sec:simulate}, it is possible to obtain $k$ realizations of $Z(t)$ conditional on $Z(0)=i$. 
The proportion of these realizations which equal $j$ can be used to approximate the transition probability $p_{i,j}(t)$. 
That is, 
\[
p_{i,j}(t) \approx \frac{1}{k}\sum_{n=1}^k 1_{\{\hat Z_n(t)=j\}}
\]
where $\hat Z_n(t)$ is a simulated realization of $Z(t)$ with $Z(0)=i$. 

We conclude this section with Table~\ref{tab:probs} which summarizes the methods for computing (or approximating) the transition probabilities and gives the label used to call them in \pkg{BirDePy}, as described in Section~\ref{sec:software}. 

\begin{table}[h!]
  \begin{center}
    \begin{tabular}{p{5.4cm}|c|p{7cm}} 
      \parbox[t]{5cm}{\textbf{Method}\\ \textbf{(in order of appearance)}} & \textbf{Label} & \textbf{Brief description}\\
      \hline
      Matrix exponential & \code{"expm"} & Uses $P(t)=\exp(Qt)$.\\
     Uniformization & \code{"uniform"} & Evaluates probability using an approximating discrete-time process. \\
      Erlangization & \code{"Erlang"} & Evaluates probability at an Erlang-distributed time. \\
     Inverse Laplace transform & \code{"ilt"} & Numerically inverts Laplace transform.\\
      Diffusion approx. & \code{"da"} & Approx.~true model by a general diffusion-scaled model. \\
     Ornstein--Uhlenbeck approx. & \code{"oua"} & Approx.\ true model by a simple diffusion process.\\
     Galton--Watson approximation & \code{"gwa"} & Approx.~true model with a linear model.\\
     Saddlepoint approximation & \code{"gwasa"} & As above combined with a saddlepoint approx.\\
     Simulation & \code{"sim"} & Average of Monte Carlo samples.\\
    \end{tabular}
        \caption{Methods for computing transition probabilities.}
    \label{tab:probs}
  \end{center}
\end{table}

\subsection{Estimation}\label{sec:estimate}
In this section we outline the parameter estimation methods implemented in \pkg{BirDePy}. 

\subsubsection{Direct numerical maximization for continuously-observed data}
For a continuously observed PSDBDP, $m$ independent `observations', or sample paths of the form $(Z(t),~t\in[0,T_k])$, are assumed to be available. 
A convenient representation for observation $k$ under this assumption is a list of jump times $t_{0,k} < t_{1,k} < \dots < t_{n_k,k}$ ($1\le k \le m$ and $n_k\ge1$), together with a corresponding list of states at those times $(z_{0,k}, z_{1,k},\dots, z_{n_k,k})$. 
Under this representation, the data, denoted by $\boldsymbol y$, consists of two lists. 
The first list contains the lists of jump times, and the second list contains the corresponding lists of states. 
The log-likelihood function for continuously-observed PSDBDPs is given by
\begin{equation}\label{eq:continuous_likelihood}
  \tilde\ell(\boldsymbol y ; \boldsymbol \theta) = \sum_{z=0}^\infty U_z\log(\lambda_z(\boldsymbol\theta))+D_z\log(\mu_z(\boldsymbol\theta)) - (\lambda_z(\boldsymbol\theta) + \mu_z(\boldsymbol\theta))H_z,
\end{equation}
where $U_z=\sum_k U^{k}_z$, $D_z=\sum_k D^{k}_z$, and $H_z=\sum_k H^{k}_z$, with $U^{k}_z$ and $ D^{k}_z$ being, respectively, the number of births and deaths whilst in state $z$ for sample path $k$, and $H^k_z$ is the total cumulative time spent in state $z$ in the time interval $[0,T_k]$ of sample path $k$. 
Therefore $U_z$, $D_z$, and $H_z$ are sufficient statistics for the data. 
For PSDBPs where the birth and death rates can be written in the form $\lambda_z = f(z)\lambda$ and $\mu_z=g(z)\mu$, where $f(z)$ and $g(z)$ are known, 
\begin{align*}
\lambda^\star = \frac{\sum_z U_z}{\sum_z f(z)H_z}\quad\text{and}\quad\mu^\star = \frac{\sum_z D_z}{\sum_z g(z)H_z}
\end{align*}
solve the maximization problem  
\[
(\lambda^\star ,\mu^\star)=:\bt^\star= \text{arg\,max}_{\bt}~\tilde\ell(\boldsymbol y ; \boldsymbol \theta).  
\]
This is discussed at length by \cite{wolff1965problems} and \cite{reynolds1973estimating}. 
These {\em maximum likelihood estimates} are the parameters that make the data most likely under the assumed model. 
For more general forms of the birth and death rates (or when the parameters defining the functions $f$ or $g$ are unknown), it is possible to find estimates $\bt^\star$ of $\bt$ by {\em directly numerically maximizing}~\eqref{eq:continuous_likelihood}. 

\subsubsection{Direct numerical maximization for discretely-observed data}
Discrete observation schemes have been receiving increasing attention in recent years.
Under these schemes, $m$ independent sample paths of $Z$ are observed, where states $z_{0,k}, z_{1,k},$$\dots, z_{n_k,k}$ of sample path $k$ are observed at respective times $t_{0,k} < t_{1,k} < \dots < t_{n_k,k}$.
In this case, the data, denoted $\boldsymbol z$, still consists of two lists. 
The first list contains the $m$ lists of observation times, and the second list contains the $m$ lists of corresponding states. 
Since the full trajectory of the process is now unobserved between observation times, $\boldsymbol z$ is different from the data $\boldsymbol y$ defined in the previous section. 

The likelihood of the discrete-observation data $\boldsymbol z$ is given by
\begin{equation}\label{eq:discrete_likelihood}
\ell(\boldsymbol z; \boldsymbol \theta) = \prod_{k=1}^m\prod_{i=1}^{n_k} p_{z_{i-1,k}, z_{i,k}}(\Delta_{i,k}),
\end{equation}where $\Delta_{i,k} := t_{i,k}-t_{i-1,k}$ denote the inter-observation times. 
Since there are generally no explicit expressions for the transition probabilities $p_{i,j}(t)$, the likelihood in \eqref{eq:discrete_likelihood}
can be approximated using the methods for approximating $p_{i,j}(t)$ discussed in Section~\ref{sec:probability}. 
Direct numerical maximization of this function then leads to approximate maximum likelihood estimates 
\[
\bt^\star = \text{arg\,max}_{\bt}~\ell(\boldsymbol z ; \boldsymbol \theta).  
\]

\subsubsection{Expectation maximization}
Another way to find maximum likelihood estimates is to treat each unobserved path of $Z$ between times $t_{i-1,k}$ and $t_{i,k}$ as `missing data', and use the EM algorithm of \cite{dempster1977maximum}. 
The EM algorithm is an iterative procedure where each iteration consists of an `E-step' and an `M-step'.  
The `E-step' is concerned with the computation of the expected value of the log-likelihood function for continuously-observed PSDBDPs \eqref{eq:continuous_likelihood}, conditional on the current iteration's parameter estimate $\bt$ and on the discretely-observed data $\boldsymbol z$.
In the `M-step', this expectation is then maximized to obtain a new parameter estimate. 

Recall the log-likelihood function for continuously observed PSDBDPs \eqref{eq:continuous_likelihood} and the fact that it depends only on the sufficient statistics $U_z$, $D_z$ and $H_z$. 
Given a pair of observations $Z(0)=i$ and $Z(t)=j$ of a birth-and-death process, it is well known (e.g., \cite{lange1995gradient,holmes2002expectation,bladt2005statistical}) that 
\begin{equation}\label{eq:em_integrals}
\begin{split}
  u_{z;i;j}(t; \bt) &:= \mathbb E[U_z \mid Z(0) = i, Z(t) = j] = \frac{\int_0^t~p_{i,z}(s)~\lambda_z~p_{z+1,j}(t-s) \di s}{p_{i,j}(t)},\\
  d_{z;i;j}(t; \bt) &:= \mathbb E[D_z \mid Z(0) = i, Z(t) = j] = \frac{\int_0^{t}~p_{i,z}(s)~\mu_z~p_{z-1,j}(t-s)\di s}{p_{i,j}(t)},\\
  h_{z;i;j}(t; \bt) &:= \mathbb E[H_z \mid Z(0) = i, Z(t) = j]  = \frac{\int_0^{t}~p_{i,z}(s)~p_{z,j}(t-s)\di s}{p_{i,j}(t)},
  \end{split}
\end{equation}
where $\lambda_z$, $\mu_z$ and $p_{i,j}(t)$ implicitly depend on $\bt$. 
Using \eqref{eq:continuous_likelihood} and \eqref{eq:em_integrals}, the conditional expected value of the log-likelihood function for continuously-observed PSDBDPs is 
\begin{equation}\label{eq:surr_likelihood}
\begin{split}
f(\boldsymbol \theta',~\boldsymbol \theta) &= \mathbb E[\tilde\ell(\boldsymbol y;\boldsymbol \theta')\mid\boldsymbol z, \boldsymbol \theta]\\
 &= \sum_{k=1}^m\sum_{i=1}^{n_k}\sum_{z=0}^\infty \Big\{u_{z;z_{i-1,k};z_{i,k}}(\Delta_{i,k}; \boldsymbol\theta)\log(\lambda_z(\boldsymbol\theta'))\\
 &\hspace{25mm}+ d_{z;z_{i-1,k};z_{i,k}}(\Delta_{i,k};\boldsymbol\theta)\log(\mu_z(\boldsymbol\theta'))\\
 &\hspace{25mm}-\big(\lambda_z(\bt')+\mu_z(\boldsymbol\theta')\big)h_{z;z_{i-1,k};z_{i,k}}(\Delta_{i,k};\boldsymbol\theta) \Big\}.
 \end{split}
\end{equation}
Starting with an arbitrary vector $\boldsymbol \theta^{(0)}$, an EM algorithm estimate follows from repeating $\boldsymbol \theta^{(k+1)} = \text{arg\,max}_{\boldsymbol\theta'} f(\boldsymbol \theta',~\boldsymbol \theta^{(k)})$, until $|\boldsymbol \theta^{(k+1)}-\boldsymbol \theta^{(k)}|$ is small or a maximum number of iterations has taken place. 

Clearly, \eqref{eq:em_integrals} needs to be computed in order to use the EM algorithm just described. 
A first possible way to do this is to simply use any of the methods from Section~\ref{sec:probability} to evaluate the transition probability functions $p_{i,j}(t)$, and then compute the integrals numerically using, for example, the trapezoidal rule. 
Upon choosing to compute $p_{i,j}(t)$ using the matrix exponential method or the inverse Laplace transform method, more sophisticated approaches are available, as we briefly describe in the next two paragraphs.

Recalling $P(t)= \exp(Qt)$ (which requires numerical approximation) from Section~\ref{sec:probability}, upon defining $\be_i$ as the $i$-th basis vector in $(|\mathcal S|+1)$-dimensional euclidean space, for any $a,b,i,j\in\mathcal S$, we can write
\[
  \int_0^t p_{i,a}(s) ~p_{b,j}(t-s) \di s = \be_{i} \int_0^t \exp(Qs) ~\be_{a}^\top\be_{b}~ \exp(Q(t-s)) \di s\, \be_{j}^\top.
\]
Let $G(t) = \int_0^t \exp(Qs) ~\be_{a}^\top\be_{b}~ \exp(Q(t-s)) \di s$. 
According to \cite{van1978computing}, if 
\[
  C := \begin{bmatrix}
      Q & \be_{a}^\top\be_{b}\\
      0 & Q
    \end{bmatrix},
\quad\text{then}\quad
  \exp(Ct) = \begin{bmatrix}
          \cdot & G(t)\\
          \cdot & \cdot
        \end{bmatrix},
\]
where `$\cdot$' denotes parts of the matrix that can be discarded.  
Hence the integrals in \eqref{eq:em_integrals} can be computed by extracting the relevant parts of the matrix exponential $\exp(Ct)$.

On the other hand, if $\mathcal L^{-1}$ denotes the inverse of a Laplace transform $\mathcal L$ (e.g., \eqref{eq:laplace}), then we also have
\[
\int_0^{t}~p_{i,a}(s)~p_{b,j}(t-s) \di s = \mathcal L^{-1}[f_{i,a}(s)~f_{b,j}(s)](t)
\]
by the properties of Laplace transforms. 
Numerical Laplace transform inversion therefore provides a third technique for evaluating the integrals in \eqref{eq:em_integrals}. 
This approach is discussed by \cite{crawford2014estimation}. 

The EM algorithm can exhibit slow convergence, calling for many iterations before a suitably accurate estimate is provided.  
Each iteration may itself use substantial computational time and effort due to requiring many matrix exponential computations, Laplace transform inversions, or numerical integration computations. 
To mitigate this, several schemes have been developed to accelerate the convergence rate of EM algorithms. 
In \citep{jamshidian1997acceleration}, four such schemes are described in detail. 
Other seminal work includes \citep{jamshidian1993conjugate} and \citep{lange1995quasi}. 
Using optimization-based ideas, these accelerators can yield substantial improvements in computational speed. 
Recall $f$ as defined in \eqref{eq:surr_likelihood} and let $\tilde {\boldsymbol g}(\boldsymbol\theta) = \big[\text{argmax}_{\boldsymbol\theta'} f(\boldsymbol \theta', \boldsymbol \theta)\big] - \boldsymbol \theta$, which is the change in parameter values when a non-accelerated EM iteration is performed. 
Loosely speaking, EM acceleration techniques treat $\tilde {\boldsymbol g}$ as a generalized derivative of the likelihood function \eqref{eq:discrete_likelihood} and aim to find $\boldsymbol \theta^\star$ such that $\tilde {\boldsymbol g}(\boldsymbol\theta^\star) = \boldsymbol 0$.

\subsubsection{Least squares estimation}
The two likelihood approaches discussed so far may be computationally demanding. 
A potential avenue to reducing the computational complexity is to instead use a least-squares-based estimator. 
Let $m_{i}(t) := \mathbb E[Z(t) \,|\, Z(0) = i]$ correspond to the 
expected population size after $t$ units of time in a birth-and-death process starting with $i$ individuals ($m_{i}(t)$ depends implicitly on the parameters $\bt$). 
The computation of $m_{i}(t)$ may be numerically less demanding than the computation of $p_{i,j}(t)$. 
Hence it may be practical to consider the least-squares estimator
\[
\boldsymbol\theta^\dagger = \text{arg\,min}_{\boldsymbol \theta}~\sum_{k=1}^m\sum_{i=1}^{n_k} \big(z_{i,k}-m_{z_{i-1,k}}(\Delta_{i,k})\big)^2.
\] 
Indeed, for linear BDPs, $m_{i}(t) = i\exp((\lambda-\mu)t)$, which is straightforward to compute, and allows a least-squares estimate to be found in this case. 
Using the Galton--Watson approach discussed in Section~\ref{sec:probability}, this can be extended to PSDBDPs. 
In addition, the deterministic approximation to the mean underlying the diffusion approximation discussed in Section~\ref{sec:probability} can also play the role of an approximate mean. 
This would certainly be less computationally demanding than calling upon the full diffusion approximation. 
Finally, any of the methods for computing approximate transition probabilities discussed in Section~\ref{sec:probability} can be used to compute an approximate expected value $m_{i}(t) = \sum_{z}p_{i,z}(t)z$. 

\subsubsection{Approximate Bayesian computation (ABC)}
The methods discussed so far have relied on using approximations to model properties which have no explicit expression (specifically, the likelihood of the observed data, and the expectation of the population size). 
ABC bypasses the need to use these model properties in the first place. 
This method works by repeatedly comparing simulated data with the observed data to find an approximation to a distribution that characterizes uncertainty about the value of $\bt$ (called the posterior distribution). 
Given a distance measure $d$, the standard ABC method (as proposed in \cite{pritchard1999population}) consists of repeating the following steps:
\begin{enumerate}
\item Generate a parameter proposal $\boldsymbol \theta$ from the prior distribution $\pi$.
\item Simulate observations $\hat{\boldsymbol z}(\boldsymbol \theta)$ consisting of points $\hat{z}_{i,k}$ generated using $\boldsymbol\theta$ with initial conditions $z_{i-1,k}$ and elapsed times $\Delta_{i,k}$ (for $k=1,\ldots,m$, $i=1,\ldots,n_k$).
\item Accept $\boldsymbol \theta$ as an approximate observation from the posterior distribution if $d\big(\boldsymbol z, \hat{\boldsymbol z}(\boldsymbol \theta)\big) < \epsilon,$ where $\boldsymbol z$ is the observed data and $\epsilon$ is a predetermined error tolerance threshold.
\end{enumerate} 
These steps are repeated until an arbitrarily large number of parameter proposals are accepted. 
A parameter point estimate can then be obtained by taking the mean or median of the accepted parameter proposals. 
While choices of $d$ and $\pi$  have been investigated for continuously-observed linear BDPs (e.g., \cite{janzen2015approximate}), 
to the authors' knowledge, there is a lack of research on suitable choices of $d$ and $\pi$ for general discretely-observed birth-and-death processes. 
A potentially suitable choice, previously discussed in the context of linear birth-and-death processes (see \cite{tavare2018linear}) is 
\begin{equation}\label{eq:ABCdistance}
  d(\boldsymbol z, \hat{\boldsymbol z}) = \sqrt{\sum_{k=1}^m\sum_{i=1}^{n_k} \big(z_{i,k}-\hat{z}_{i,k}(\boldsymbol \theta)\big)^2}. 
\end{equation}
The prior $\pi$ can simply be taken as a uniform distribution on the set of allowed parameters.  

The choice of threshold $\epsilon$ in Step~3 above has a strong influence on the probability of a parameter proposal being accepted. 
Larger choices of $\epsilon$ may allow the desired number of accepted parameter proposals to be found more quickly, but this could be at the expense of accuracy. 
Similarly, the choice of the prior distribution $\pi$ has a major impact on performance. 
To address these concerns, the standard ABC method can be applied iteratively with dynamically determined $\epsilon$ and $\pi$. 
Early work in this direction can be found in \cite{beaumont2009adaptive} and \cite{del2012adaptive}. 
More recently, \cite{simola2021adaptive} proposed a method for adaptively selecting a threshold at each iteration by comparing the estimated posterior from the two previous iterations and the distances between the sampled data and the accepted parameter proposals in the previous iteration.

\subsection{Forecasting}\label{sec:forecast}
Confidence and prediction intervals are two simple ways to convey information about possible future states of a stochastic process. 
Confidence intervals are focused on parameter uncertainty, while prediction intervals incorporate parameter uncertainty and model stochasticity. 
Let $Z_{\bt}$ be a birth-and-death process evolving according to parameters $\bt$. 
Given $Z(0)=i$, associated with each possible $\bt$ is an expected trajectory of the process $\big(m_{\bt}(t), t\in[0,T]\big)$, where $m_{\bt}(t) = \mathbb E Z_{\bt}(t)$. 
The paths $m_{\bt}$ can, for example, be approximated using simulation or by the mean curve of the diffusion approximation described in Section~\ref{sec:probability}. 
When uncertainty about parameter values is characterized by some distribution, a {\em confidence interval} can be formed by sampling from this distribution (e.g., the distribution of an estimator
or the posterior distribution) and collecting together the associated approximations of $m_{\bt}$. 
Alternatively, each sampled $\bt$ can be used to generate a sample $(Z_{\bt}(t), t\in[0,T])$ with $Z_{\bt}(0)=i$. 
Collecting these samples together allows for an approximate {\em prediction interval} to be formed. 

\section[BirDePy description]{\pkg{BirDePy} description} \label{sec:software}
In this section we describe how to use \pkg{BirDePy}. 
As is usual for Python we use a shorthand \code{bd} in place of the full package name \code{birdepy}. 
The package has five core functions: \code{bd.simulate.continuous()}, \code{bd.simulate.discrete()}, \code{bd.probability()}, \code{bd.estimate()} and \code{bd.forecast()}.
In addition the two functions \code{bd.gpu_functions.probability()} and \code{bd.gpu_functions.discrete()} are available when the system has a CUDA-enabled GPU with compute capability 2.0 or above with an up-to-date Nvidia driver. 
These CUDA based functions provide a limited version of the functionality in \code{bd.probability()} and \code{bd.simulate.discrete()} but are often capable of providing output substantially faster. 

\pkg{BirDePy} can be installed by opening Python (v3.7 or greater) and executing \code{pip install birdepy}. 
It can then be imported to a session using \code{import birdepy as bd}. 

In order to use the core functions of \pkg{BirDePy} the packages \pkg{NumPy} ($\ge$v1.17.0) (see \cite{harris2020array}), \pkg{mpmath}(v1.1.0 or greater) (see \cite{mpmath}), \pkg{SciPy} (v1.7.0 or greater) (see \cite{virtanen2020scipy}), \pkg{matplotlib} (v3.4.2 or greater) (see \cite{hunter2007matplotlib}) and \pkg{gwr-inversion} need to be installed. 
The CUDA based functions contained in the module \code{bd.gpu_functions} additionally require \pkg{Numba} (v0.53.1 or greater) (see \cite{lam2015numba}) and \pkg{cudatoolkit} (v9.2 or greater) to be installed. 

In all instances parameters are passed to functions as a {\em list} in the canonical order they are listed in Table~\ref{tab:rates}.  

\subsection[Continuous simulation: bd.simulate.continuous()]{Continuous simulation: \code{bd.simulate.continuous()}} \label{sec:bd.simulate.continuous}
This function is used to simulate PSDBDPs at birth and death event times.

\subsubsection{Usage and default argument values}
\begin{CodeInput}
bd.simulate.continuous(param, model, z0, t_max, k=1, survival=False, 
                       seed=None, **options)
\end{CodeInput}
\subsubsection{Arguments}

\code{param} {\em (array like)} ---  The parameters governing the evolution of the birth-and-death process to be simulated. Array of real elements with size \code{m}, where \code{m} is the number of parameters.  

\code{model} {\em (str)} --- Choice of model as described in Section~\ref{sec:models}. Should be one of the choices in Table~\ref{tab:rates} or \code{"custom"} (see Section~\ref{sec:custom} for details on specifying custom models).  

\code{z0} {\em (int or callable)} --- The initial population size for each sample path. If it is a callable it should be a function that has no arguments and returns an integer, for example if the initial population follows a geometric distribution with parameter $0.35$, then \code{z0} could be specified by \code{z0 = lambda: np.random.default_rng().geometric(0.35)}. 

\code{t_max} {\em (scalar)} --- The simulation horizon. All events up to and including this time are included in the output.

\code{k} {\em (int, optional)} --- The number of independent sample paths to be simulated.

\code{survival} {\em (bool, optional)} ---  If set to \code{True}, then the simulated sample paths are conditioned to have a positive population size at the final observation time. Since this uses acceptance-rejection it can greatly increase computation time. 

\code{seed} {\em (int, Generator, optional)} ---  If \code{seed} is not specified the random numbers are generated according to \code{numpy.random.default_rng()}. If \code{seed} is an integer, random numbers are generated according to \code{numpy.random.default_rng(seed)}. If seed is a \code{Generator}, then that object is used. 

\subsubsection{Returns}

\code{jump_times} {\em (list)} ---  If \code{k} is \code{1}, then this records a list containing jump times, generated according to \code{model}. If \code{k} is greater than 1, then this records a list of lists where each list corresponds to the jump times from one sample path.

\code{pop_sizes} {\em (list)} ---  If \code{k} is 1, then this records a list containing population sizes at the corresponding elements of \code{jump_times}, generated according to \code{model}. If \code{k} is greater than 1, then this records a list of lists where each list corresponds to the population sizes corresponding to \code{jump_times} from one sample path.

\subsection[Discrete simulation: bd.simulate.discrete()]{Discrete simulation: \code{bd.simulate.discrete()}} \label{sec:bd.simulate.discrete}
This function is used to simulate PSDBDPs at discrete observation times.

\subsubsection{Usage and default argument values}
\begin{CodeInput}
bd.simulate.discrete(param, model, z0, times, k=1, method="exact", tau=0.1,
                     survival=False, seed=None, display=False, **options)
\end{CodeInput}
\subsubsection{Arguments}
This function has all of the parameters listed for \code{bd.simulate.continuous()} described in Section~\ref{sec:bd.simulate.continuous} except for parameter \code{t_max}. 
In addition it has parameters: 

\code{times} {\em (array like)} --- The discrete times at which the simulated birth-and-death is observed. Array of real elements of size \code{n}, where \code{n} is the number of observation times.

\code{method} {\em (str, optional)} --- Simulation algorithm used to generate samples as described in Section~\ref{sec:simulate}. Should be one of the choices in Table~\ref{tab:sims}.  

\code{tau} {\em (scalar, optional)} ---  Time between samples for the approximation methods \code{"ea"}, \code{"ma"} and \code{"gwa"} described in Table~\ref{tab:sims}.

\code{display} {\em (bool, optional)} ---  If set to \code{True}, then a progress indicator is printed as the simulation is performed.

\subsubsection{Returns}

\code{out} {\em (array like)} ---  If \code{k} is equal to 1, then this records a list containing sampled population size observations at \code{times}, generated according to \code{model}. Or if \code{k} is greater than 1, then a {\em numpy.ndarray} containing \code{k} sample paths, each contained in a row of the array, is generated. 

\subsection[Discrete simulation on GPU: bd.gpu functions.discrete()]{Discrete simulation on GPU: \code{bd.gpu\_functions.discrete()}} \label{sec:bdg.discrete}
Exact simulation of continuous-time birth-and-death processes at a specified time using CUDA.

\subsubsection{Usage and default argument values}
This function is not imported by default with \pkg{BirDePy}, the \code{gpu_functions} module needs to be imported: 
\begin{CodeInput}
import birdepy.gpu_functions as bdg
bdg.discrete(param, model, z0, t, k=1, survival=False, seed=1)
\end{CodeInput}
\subsubsection{Arguments}
This function has almost the same parameters as \code{bd.simulate.discrete()} described in Section~\ref{sec:bd.simulate.discrete}. 
It does not have the parameters \code{method}, \code{tau} or \code{display}, and parameter \code{t} replaces parameter \code{time}. 
It does not accept custom model functionality. 

\code{t} {\em (scalar)} ---  The time at which the simulated birth-and-death is observed.

\subsubsection{Returns}

\code{out} {\em (array like)} ---  A list containing \code{k} sampled population size observations at time \code{t} which are generated according to \code{model}.

\subsection[Transition probabilities: bd.probability()]{Transition probabilities:  \code{bd.probability()}} \label{sec:bd.probability}
This function computes approximations to transition probabilities of PSDBDPs. 

\subsubsection{Usage and default argument values}

\code{bd.probability(z0, zt, t, param, model="Verhulst", method="expm", **options)}

\subsubsection{Arguments}

\code{z0} {\em (array like)} --- States of birth-and-death process at time $0$ 

\code{zt} {\em (array like)} --- States of birth-and-death process at time(s) $t$ 

\code{t} {\em (array like)} --- Elapsed time(s) (if this has size greater than 1, then it must be increasing)

\code{param} {\em (array like)} --- The parameters governing the evolution of the birth-and-death process. Array of real elements of size \code{n}, where \code{n} is the number of parameters.  

\code{model} {\em (str, optional)} --- Model specifying birth and death rates of process as described in Section~\ref{sec:models}. Should be one of the choices in Table~\ref{tab:rates} or \code{"custom"} (see Section~\ref{sec:custom} for details on specifying custom models).  

\code{method} {\em (str, optional)} --- Transition probability approximation method as described in Section~\ref{sec:probability}. Should be one of the choices in Table~\ref{tab:probs}.  

\code{options} {\em (dict, optional)} --- A dictionary of method specific options.  
Methods \code{"uniform"}, \code{"Erlang"}, \code{"ilt"}, \code{"da"} enable an optional parameter \code{k} which tunes accuracy. 
Methods \code{"expm"}, \code{"uniform"}, \code{"Erlang"} enable an optional parameter \code{z_trunc} which dictates truncation thresholds, i.e., minimum and maximum states of process considered; this is an array of real elements of size 2, by default \code{z_trunc=[z_min, z_max]} where \code{z_min=max(0, min(z0, zt) - 100)} and \code{z_max=max(z0, zt) + 100}. 
Method \code{"ilt"} also enables optional parameters: (i) \code{lent_eps} which is a {\em scalar} termination threshold for the Lentz algorithm computation of Equation~\eqref{eq:laplace_ij}, (ii) \code{laplace_method} which is a {\em str} that determines which Laplace inversion method from Table~\ref{tab:laplace} to use, and (iii) \code{precision} which is an integer that determines the numerical precision to use. 

\begin{table}[h!]
  \begin{center}
    \begin{tabular}{p{4.5cm}|p{9cm}} 
      \parbox[t]{4.5cm}{\textbf{Method label}} & \textbf{Brief description/reference}\\
      \hline
      \code{"cme-talbot"} (default) & Attempts method \code{"cme-mp"} and if an error occurs attempts method \code{"talbot"}. \\
     \code{"cme"} & `CME" method of \cite{horvath2020numerical}, implementation downloaded from \cite{inverselaplace_website}. \\
      \code{"euler"} & `Euler' method of \cite{abate2006unified}, implementation downloaded from \cite{inverselaplace_website}. \\
     \code{"gaver"} & Function \code{mpmath.invertlaplace(method="gaver")} of \cite{mpmath}.\\
     \code{"talbot"} & Function \code{mpmath.invertlaplace(method="talbot")} of \cite{mpmath}.\\
     \code{"stehfest"} & Function \code{mpmath.invertlaplace(method="stehfest")} of \cite{mpmath}.\\
     \code{"dehoog"} & Function \code{mpmath.invertlaplace(method="dehoog")} of \cite{mpmath}.\\
     \code{"cme-mp"} & Version of \code{"cme"} coded using \pkg{mpmath} operations. \\
     \code{"gwr"} & Method of \cite{valko2004comparison} as implemented in \code{gwr_inversion.gwr()}.\\
    \end{tabular}
        \caption{Methods for performing numerical Laplace transform inversion available in \pkg{BirDePy}.}
    \label{tab:laplace}
  \end{center}
\end{table}

\subsubsection{Returns}

\code{transition_probability} {\em (numpy.ndarray)} --- An array of transition probabilities. If \code{t} has size bigger than $1$, then the first coordinate corresponds to \code{t}, the second coordinate corresponds to \code{z0} and the third coordinate corresponds to \code{zt}; for example if \code{z0=[1,3,5,10]}, \code{zt=[5,8]} and \code{t=[1,2,3]}, then \code{transition_probability[2,0,1]} corresponds to $\mathbb P(Z(3)=8\mid Z(0)=1)$. If \code{t} has size $1$, then the first coordinate corresponds to \code{z0} and the second coordinate corresponds to \code{zt}.\\

\subsection[Transition probabilities on GPU: bd.gpu functions.probability()]{Transition probabilities on GPU: \code{bd.gpu\_functions.probability()}} \label{sec:bd.probabilitygpu}
This function computes transition probabilities for PSDBDPs using Monte Carlo simulation on a GPU.

\subsubsection{Usage and default argument values}
This function is not imported by default with \pkg{BirDePy}, the \code{gpu_functions} module needs to be imported: 

\code{import birdepy.gpu_functions as bdg\\
bdg.probability(z0, zt, t, param, model, k=10**6, seed=1)}

\subsubsection{Arguments}
This function has almost the same parameters as \code{bd.probability()}, described in Section~\ref{sec:bd.probability}. 
It does not have the parameters \code{method} or \code{options}. 
In addition it does have:

\code{k} {\em (int, optional)} ---  Minimum number of samples used to generate each probability estimate (actual number of samples will usually be higher due to the way memory is allocated on GPU). The total number of samples used will be at least \code{z0.size * k}.

\subsubsection{Returns}
As for \code{bd.probability()} described in Section~\ref{sec:bd.probability}.

\subsection[Parameter estimation: bd.estimate()]{Parameter estimation:  \code{bd.estimate()}} \label{sec:bd.estimate}
This function performs parameter estimation for (continuously or discretely observed) PSDBDPs. 
Depending on the argument of \code{framework}, various optional input parameters become available; these optional input parameters are described in Section~\ref{sec:bd.estimate_abc}, Section~\ref{sec:bd.estimate_dnm}, Section~\ref{sec:bd.estimate_em} and Section~\ref{sec:bd.estimate_lse}. 

\subsubsection{Usage and default argument values}
\begin{CodeInput}
bd.estimate(t_data, p_data, p0, p_bounds, framework="dnm", model="Verhulst",
            scheme="discrete", con=(), known_p=(), idx_known_p=(), 
            se_type="asymptotic", ci_plot=False, export=False, display=False, 
            **options)
\end{CodeInput}
\subsubsection{Arguments}

\code{t_data } {\em (list)} --- Observation times of birth-and-death process. If one trajectory is observed, then this is a list. If multiple trajectories are observed, then this is a list of lists where each list corresponds to a trajectory.

\code{p_data } {\em (list)} --- Observed population sizes of birth-and-death process at times in parameter \code{t_data}. If one trajectory is observed, then this is a list. If multiple trajectories are observed, then this is a list of lists where each list corresponds to a trajectory.

\code{p0} {\em (array like)} --- Initial parameter guess. Array of real elements of size \code{m}, where \code{m} is the number of unknown parameters.

\code{p_bounds} {\em (list)} ---  Bounds on parameters. Should be specified as a sequence of (min, max) pairs for each unknown parameter. See Section~\ref{sec:bounds} for more information. 

\code{framework} {\em (str, optional)} --- Parameter estimation framework. Should be one of: (i) \code{"abc"} (see Section~\ref{sec:bd.estimate_abc}), (ii) \code{"dnm"} (see Section~\ref{sec:bd.estimate_dnm}), (iii) \code{"em"} (see Section~\ref{sec:bd.estimate_em}), or (iv) \code{"lse"} (see Section~\ref{sec:bd.estimate_lse}). 

\code{model} {\em (str, optional)} --- Model specifying birth and death rates of process as described in Section~\ref{sec:models}. Should be one of the choices in Table~\ref{tab:rates} or \code{"custom"} (see Section~\ref{sec:custom} for details on specifying custom models).  

\code{scheme} {\em (str, optional)} --- Observation scheme. Should be one of: (i) \code{"discrete"} or (ii) \code{"continuous"}.  
If set to \code{"continuous"}, then it is assumed that the population is observed continuously with jumps occurring at times in \code{t_data} into corresponding states in \code{p_data}.

\code{con} {\em (\{Constraint, dict\} or List of \{Constraint, dict\}, optional)} ---   Constraints definition for parameters.  See Section~\ref{sec:bounds} for more information. 

\code{known_p} {\em (array like, optional)} ---  List of known parameter values. For built in models these must be in their canonical order as given in Table~\ref{tab:rates}. If this argument is given, then \code{idx_known_p} must also be specified.  See Section~\ref{sec:bounds} for more information. 

\code{idx_known_p} {\em (array like, optional)} ---  List of indices of known parameters (as given in argument \code{known_p}). For built in models indices must correspond to canonical order as given in Table~\ref{tab:rates}. If this argument is given, then argument \code{known_p} must also be specified.  See Section~\ref{sec:bounds} for more information. 

\code{se_type} {\em (str, optional)} ---  Should be one of: (i) \code{"none"}, (ii) \code{"simulated"}, or (iii) \code{"asymptotic"}. See Section~\ref{sec:confidence} for more information.

\code{ci_plot} {\em (bool, optional)} ---  Enables confidence region plotting for 2-dimensional parameter estimates. See Section~\ref{sec:confidence} for more information.

\code{export} {\em (str, optional)} --- File name for export of confidence region figure to a \LaTeX~file.

\code{display} {\em (bool, optional)} ---  If set to \code{True}, then a progress indicator is printed for some methods.

\subsubsection{Returns}
An object \code{res} with the following attributes is returned. 
Note: not all attributes are relevant to all frameworks and schemes. 

\code{p} {\em (list)} ---  Parameter estimate. 

\code{capacity} {\em (list)} ---  Estimated possible carrying capacity values. 

\code{val} {\em (scalar)} ---  If \code{framework} is \code{"dnm"} or \code{"em"}: value of log-likelihood at parameter estimate. If \code{framework} is \code{"lse"}: value of squared error at parameter estimate. 

\code{cov} {\em (array like)} ---  Covariance matrix describing uncertainty of parameter estimates. 

\code{se} {\em (list)} ---  Standard errors of parameter estimates. 

\code{compute_time} {\em (scalar)} ---  Total CPU time used by function (seconds). 

\code{framework} {\em (str)} ---  Value of \code{framework} argument input. 

\code{message} {\em (str)} ---  Message from optimization routine (if used). 

\code{success} {\em (bool)} ---  Success indicator from optimization routine (if used). 

\code{iterations} {\em (list)} ---  Parameter estimates at each iteration. Relevant to ABC and EM frameworks. 

\code{method} {\em (str)} ---  Value of \code{method} argument input. 

\code{p0} {\em (list)} ---  Initial parameter value. 

\code{scheme} {\em (str)} ---  Value of \code{scheme} argument input. 

\code{samples} {\em (list)} ---  Accepted samples for the ABC framework.

\subsection[ABC algorithm: bd.estimate(framework="abc")]{ABC algorithm: \code{bd.estimate(framework="abc")}} \label{sec:bd.estimate_abc}
This function performs parameter estimation for discretely-observed PSDBDPs using an ABC algorithm. 
This framework is called using the function \code{bd.estimate()} described in Section~\ref{sec:bd.estimate}. 

\subsubsection{Usage and default argument values}
\begin{CodeInput}
bd.estimate(t_data, p_data, p0, p_bounds, framework='abc', eps_abc='dynamic', 
            k=100, max_its=3, max_q=0.99, eps_change=5, gam=5, method='gwa', 
            tau=None, seed=None, distance=None, stat='mean', display=False)
\end{CodeInput}
\subsubsection{Arguments}
In addition to the parameters of \code{bd.estimate()} described in Section~\ref{sec:bd.estimate}:

\code{eps_abc} {\em (list, str, optional)} --- Threshold error tolerance for distance between simulated data and observed data for accepting parameter proposals.  If this is set to \code{"dynamic"} (default), then a slightly modified version of the procedure described in \citep{simola2021adaptive} is used. Otherwise \code{eps_abc} must be a list which specifies the threshold for each iteration. 

\code{k} {\em (int, optional)} --- Number of successful parameter samples used to obtain the estimate. 

\code{max_its} {\em (int, optional)} --- Maximum number of iterations of the algorithm.

\code{max_q} {\em (scalar, optional)} --- Tolerance threshold for stopping algorithm (see Equation (2.5) in \citep{simola2021adaptive}). This is only checked after at least two iterations have occurred. Should be selected from $(0,1]$. 

\code{eps_change} {\em (scalar, optional)} --- An iteration is only performed if the percentage decrease in the tolerance threshold compared to the previous iteration is greater than this value. If \code{eps_change=0} and \code{eps_abc="dynamic"}, then the procedure described in \citep{simola2021adaptive} is used.

\code{gam} {\em (int, optional)} --- If \code{eps_abc} is set to \code{"dynamic"}, then \code{k*gam} parameter values are initially sampled and the distance between the data and the $k$-th largest distance corresponding to these samples is used as the first tolerance threshold. 

\code{distance} {\em (callable, optional)} --- Computes the distance between simulated data and observed data. The default value is \eqref{eq:ABCdistance}. 

\code{stat} {\em (str)} ---  Determines which statistic is used to summarize the posterior distribution. Should be one of: \code{"mean"} or \code{"median"}.

This function also accepts the optional parameters \code{method}, \code{tau}, \code{seed}, \code{display} as described in Section~\ref{sec:bd.simulate.discrete}. 

\subsection[DNM algorithm: bd.estimate(framework="dnm")]{DNM algorithm: \code{bd.estimate(framework="dnm")}} \label{sec:bd.estimate_dnm}
This function performs parameter estimation for discretely observed PSDBDPs using direct numerical maximization of approximate likelihood functions.
This framework is called using the function \code{bd.estimate()} described in Section~\ref{sec:bd.estimate}. 

\subsubsection{Usage and default argument values}
\begin{CodeInput}
bd.estimate(t_data, p_data, p0, p_bounds, framework="dnm", likelihood="expm",
            z_trunc=())
 \end{CodeInput}
\subsubsection{Arguments}
In addition to the parameters of \code{bd.estimate()} described in Section~\ref{sec:bd.estimate}: (i) the parameter \code{likelihood} as described in Section~\ref{sec:bd.probability} is available, and (ii) when parameter \code{likelihood} has value \code{"expm"}, \code{"uniform"} or \code{"Erlang"} the parameter \code{"z_trunc"} as described in Section~\ref{sec:bd.probability} becomes available. 

\subsection[EM algorithm: bd.estimate(framework="em")]{EM algorithm: \code{bd.estimate(framework="em")}} \label{sec:bd.estimate_em}
This function performs parameter estimation for discretely-observed PSDBDPs using an EM algorithm. 
This framework is called using the function \code{bd.estimate()} described in Section~\ref{sec:bd.estimate}. 

\subsubsection{Usage and default argument values}
\begin{CodeInput}
bd.estimate(t_data, p_data, p0, p_bounds, framework="em", technique="expm",
            accelerator="none", likelihood="expm", laplace_method="cme-talbot",
            lentz_eps=1e-6, max_it=100,i_tol=1e-3, j_tol=1e-2, h_tol=1e-2,
            z_trunc=())
\end{CodeInput}
\subsubsection{Arguments}
In addition to the parameters of \code{bd.estimate()} described in Section~\ref{sec:bd.estimate}: 

\code{technique} {\em (str, optional)} --- Determines how the integrals in Equation~\eqref{eq:em_integrals} are evaluated. Must be one of \code{"expm"}, \code{"ilt"} or \code{"num"} which respectively result in matrix exponential, inverse Laplace transform and numerical methods being called upon. 

\code{accelerator} {\em (str, optional)} --- EM accelerator to be used. Must be one of \code{"cg"}, \code{"none"}, \code{"Lange"}, \code{"qn1"} or \code{"qn2"}. The first option uses the method of \cite{jamshidian1993conjugate}, the second uses the method of \cite{dempster1977maximum}, the third uses the method of \cite{lange1995quasi}, and the final two use methods from \cite{jamshidian1997acceleration}.  

\code{i_tol} {\em (scalar, optional)} ---  Algorithm terminates when \code{sum(abs(p(i) - p(i-1)) < i_tol} where \code{p(i)} and \code{p(i-1)} are parameter estimates corresponding to iteration $i$ and $i-1$.

\code{j_tol} {\em (scalar, optional)} ---  States $z$ with expected number of upward transitions $u_{z;i;j}$ (as defined in Equation~\eqref{eq:em_integrals}) or expected number of downward transitions $d_{z;i;j}$ (as defined in Equation~\eqref{eq:em_integrals}) greater than \code{j_tol} are included in E steps. 

\code{h_tol} {\em (scalar, optional)} ---  States $z$ with expected holding time $h_{z;i;j}$ (as defined in Equation~\eqref{eq:em_integrals}) greater than \code{h_tol} are included in E steps.

As described in Section~\ref{sec:bd.probability} this function also accepts the optional parameters \code{likelihood}, \code{laplace_method}, \code{lentz_eps} and \code{z_trunc}. 

\subsection[LSE algorithm: bd.estimate(framework="lse")]{LSE algorithm: \code{bd.estimate(framework="lse")}} \label{sec:bd.estimate_lse}
This function performs parameter estimation for PSDBDPs using least squares estimation.  
This framework is called using the function \code{bd.estimate()} described in Section~\ref{sec:bd.estimate}. 

\subsubsection{Usage and default argument values}
\begin{CodeInput}
bd.estimate(t_data, p_data, p0, p_bounds, framework="lse", squares="fm",
            z_trunc=())
\end{CodeInput}

\subsubsection{Arguments}
In addition to the parameters of \code{bd.estimate()} described in Section~\ref{sec:bd.estimate}: 

\code{squares} {\em (str, optional)} --- Method used to compute approximate expected values so that the squared difference between observed data and the expected value can be approxiamted. Should be one of \code{"expm"}, \code{"fm"} or \code{"gwa"} which respectively use a matrix exponential, diffusion mean or linear approximation approach to approximate the expected value. 

When parameter \code{squares} is set to \code{"expm"} optional parameter \code{"z_trunc"} as described in Section~\ref{sec:bd.probability} becomes available. 

\subsection[Forecasting: bd.forecast()]{Forecasting: \code{bd.forecast()}} \label{sec:bd.forecast}
This function performs simulation or numerical based forecasting for PSDBDPs. 
It produces a plot of the likely range of mean population sizes subject to parameter uncertainty (confidence intervals) or the likely range of population sizes subject to parameter uncertainty and model stochasticity (prediction intervals). 

\subsubsection{Usage and default argument values}
\begin{CodeInput}
bd.forecast(model, z0, times, param, cov=None, interval="confidence", 
            method=None, 
            percentiles=(0, 2.5, 10, 25, 50, 75, 90, 97.5, 100),
            labels=('$95\%$', '$80\%$', '$50\%$'),
            p_bounds=None, con=(), known_p=(), idx_known_p=(), k=10 ** 3, 
            n=10 ** 3, seed=None, colormap=cm.Purples, xlabel="Time", 
            ylabel="default", xticks="default", rotation=45, display=False,
            export=False, **options)
\end{CodeInput}
\subsubsection{Arguments}
This function shares some parameters with \code{bd.simulate()} and \code{bd.estimate()}. 
Parameters \code{model}, \code{z0}, \code{times}, \code{param}, \code{seed} and \code{display} are described in Section~\ref{sec:bd.simulate.discrete}. 
Parameters \code{p_bounds}, \code{con}, \code{known_p}, and \code{idx_known_p} are described in Section~\ref{sec:bd.estimate}. 
In addition:

\code{interval} {\em (str, optional)} --- Type of forecast. Should be one of \code{"confidence"} (default) or \code{"prediction"}. Confidence intervals show the likely range of mean future population values, reflecting parameter uncertainty. Prediction intervals show the likely range of future population values, incorporating parameter uncertainty and model stochasticity.

\code{method} {\em (str, optional)} --- Method used to generate samples. For confidence intervals samples are trajectories of future expected values. For prediction intervals samples are trajectories of future population values. Should be one of: \code{"fm"} (default for confidence intervals), \code{"exact"}, \code{"ea"}, \code{"ea"}, \code{"ma"} or \code{"gwa"} (default for prediction intervals) . 

\code{cov} {\em (array like, optional)} ---  If this is specified, then the parameters are assumed to follow a truncated normal distribution with this covariance (and mean \code{param}). In this case parameter \code{p_bounds} should also be specified to avoid unwanted parameters. 

\code{percentiles} {\em (list, optional)} ---  List of percentiles to split data into. 

\code{labels} {\em (list, optional)} ---  List of strings containing labels for each percentile split. 

\code{k} {\em (int, optional)} --- Number of samples used to generate the forecast. For confidence intervals each sample corresponds to an estimate of the mean for a parameter value sampled from a normal distribution with mean given by \code{param} and covariance given by \code{cov}. For prediction intervals each sample corresponds to a simulated trajectory of population size for a parameter value sampled in the way just described.

\code{n} {\em (int, optional)} --- Number of samples used to estimate each sample of a mean for confidence interval samples. Only applicable when method is \code{"exact"}, \code{"ea"}, \code{"ma"} or \code{"gwa"}.

\code{colormap} {\em (matplotlib.colors.LinearSegmentedColormap, optional)} --- Colors used for plot.

\code{xlabel} {\em (str, optional)} --- Label for x axis of plot.

\code{ylabel} {\em (str, optional)} --- Label for y axis of plot.

\code{xticks} {\em (array like, optional)} --- Locations of x ticks.

\code{rotation} {\em (int, optional)} --- Rotation of x tick labels.

\code{export} {\em (boolean, optional)} --- If \code{True}, then a \LaTeX~file containing the figure in \proglang{PGF/TikZ} is generated.

\subsection{Custom models}\label{sec:custom}
It is possible to specify custom birth and death rates for the five core \pkg{BirDePy} functions. 
This is achieved by setting \code{model} to \code{"custom"} and additionally passing {\em callables} to arguments \code{b_rate} and \code{d_rate}. These callables must take as input a population size \code{z} and a list of parameters \code{p}, and respectively return scalars corresponding to the birth and death rates at population size \code{z}. 
For example: \code{b_rate = lambda z, p: p[0]*z**2} and \code{b_rate = lambda z, p: p[0]*z}.

\subsection{Known parameters, optimization options and constraints}\label{sec:bounds}
It may be the case that some of the parameters of a model are known and others are unknown. 
In this case, the parameters \code{known_p} and \code{idx_known_p} can be passed to \code{bd.estimate()} or \code{bd.forecast()}. 
The parameter \code{known_p} is an ordered list of known parameter values and \code{idx_known_p} is an ordered list containing the indices of the known parameters as they appear in the parameter \code{param} in the function \code{bd.probability()}. 
For built-in models the indices of parameters correspond to their canonical order in Table~\ref{tab:rates}.

By default, when an optimization routine is needed, \code{birdepy.estimate()} uses the option \code{"L-BFGS-B"} of \code{scipy.optimize.minimize()}, unless constraints are specified in which case \code{"SLSQP"} is used instead. 
The other choices of method in \code{scipy.optimize.minimize()} can be accessed in \code{birdepy.estimate()} using the optional parameter \code{opt_method}. 
It is also possible to use \code{scipy.optimize.differential_evolution()} by setting \code{opt_method} to \code{"differential-evolution"}.
In addition, many of the optional arguments of the functions \code{scipy.optimize.minimize()} and \code{scipy.optimize.differential_evolution()} can be used as optional parameters in \code{birdepy.estimate()}.

Constraints on parameters are passed to \code{bd.estimate()} or \code{bd.forecast()} through the parameter \code{con}. 
This is in addition to the bounds that are passed to parameter \code{p_bounds}. 
Constraints are used by the \code{scipy.optimize} functions \code{minimize()} or \code{differential_evolution()}. 
Each constraint must be a dictionary with fields:

\code{type} {\em (str)} --- Constraint type: \code{"eq"} for equality, \code{"ineq"} for inequality.

\code{fun} {\em (callable)} --- The function defining the constraint.

\code{args} {\em (sequence, optional)} --- Extra arguments to be passed to the function.

Equality constraint means that the constraint function result is to be zero, whereas inequality means that it is to be non-negative. 
Note that setting \code{opt_method} to \code{"COBYLA"} means only inequality constraints are supported.

\subsection{Standard errors and confidence regions}\label{sec:confidence}
The attributes and parameters in this section relate to the function \code{bd.estimate()} and its output \code{res}. 
The confidence regions described here are related to ---but not the same as--- the confidence intervals described in Section~\ref{sec:forecast} (which are relevant to \code{bd.forecast()} as described in Section~\ref{sec:bd.forecast}). 
Available methods for computing confidence regions and standard errors to quantify the uncertainty of parameter estimates depends on the framework being employed. 
This section is concerned with the attributes \code{cov} and \code{se} of the output of \code{bd.estimate()}, and the confidence region plot that is printed when \code{ci_plot} is set to \code{True}. 
The behavior of these objects depends on the argument of \code{se_type}, which may be set to \code{"none"}, \code{"asymptotic"} or  \code{"simulated"}.  
If \code{"none"} is chosen, then \code{cov} and \code{se} will be empty and setting \code{ci_plot} to \code{True} will result in an error. 

For frameworks \code{"dnm"} and \code{"em"}, when \code{se_type} is set to \code{"asymptotic"}, a covariance matrix is returned in attribute \code{cov} as the negative of the inverse of the Hessian matrix of the approximate log-likelihood function at the point of the parameter estimate contained in \code{res.p}. 
In this case, standard errors are returned in the attribute \code{se} by assuming a normal distribution with mean given by attribute \code{p} and covariance given by attribute \code{cov}. 
In addition, when two parameters are being estimated, setting the optional parameter \code{ci_plot} to \code{True} results in a plot of confidence regions generated according to this assumed normal distribution.

For frameworks \code{"dnm"}, \code{"em"} and \code{"lse"}, when \code{se_type} is set to \code{"simulated"}, given a parameter estimate, the following steps are repeated:
\begin{enumerate}
\item Simulate data $\hat {\boldsymbol z}$ which has the same form as $\boldsymbol z$, that is, consisting of points generated using \code{bd.simulate.discrete()} with the arguments of \code{param}, \code{z0} and \code{times} set respectively to \code{res.p}, \code{p_data[i][0]} and \code{t_data[i]} for \code{i} in \code{range(len(p_data))}. 
\item Apply the same parameter estimation technique to find the estimate corresponding to $\hat {\boldsymbol z}$. 
\end{enumerate}
These steps are repeated $100$ times (or a different number of times, as specified by parameter \code{num_samples}), and then a multivariate normal is fitted to the collection of estimates that are generated at Step~2 of each iteration. 
From this a covariance matrix is returned as the attribute \code{cov}, which characterizes the uncertainty of the estimate given by attribute \code{p}. 
In this case, this multivariate normal is used to generate: (i) standard errors which are recorded in attribute \code{se}, and (ii) the confidence region plotted when \code{ci_plot} is set to \code{True}. 

For framework \code{"abc"}, the attribute \code{samples} is returned, which is a recording of the accepted parameter proposals. 
Performing a density estimate on these samples is a standard method of characterizing uncertainty for ABC estimates. 
In addition, attribute \code{cov} is returned as an estimate of the covariance of the final iteration accepted parameter proposals, and this is also used to generate the standard errors returned in attribute \code{se}.

\section{Numerical examples}\label{sec:examples}
In this section we present several numerical examples to show \pkg{BirDePy} in action. 
The goal of these examples is to demonstrate the execution of the vast majority of the \pkg{BirDePy} codebase. 
In particular, an effort is made to ensure that functions and methods which are not used in the case study in Section~\ref{sec:case} are used in this section. 
It is out of the scope of this paper to provide comprehensive evidence to compare the alternative methods implemented within each function. 
We do, however, provide CPU times to give users a rough indication of how long each operation may be expected to take. 
Throughout the section, we assume the following commands have already been executed:
\begin{CodeInput}
import birdepy as bd
import birdepy.gpu_functions as bdg
import numpy as np
\end{CodeInput}

In this section, numbers in $[0.0001, 1)$ are rounded to $4$ decimal places, numbers in $[1,10)$ are rounded to $2$ decimal places, and numbers greater than $10$ are rounded to the nearest whole number. 
Any output which was returned as a \code{nan} value is indicated by *. 
The models used in our examples are discussed in more detail in Section~\ref{sec:models} and summarized in Table~\ref{tab:rates}. 

In Section~\ref{sec:num_sim} we demonstrate the use of the simulation functions \code{bd.simulate.discrete()}, \code{bd.simulate.continuous()} and \code{bdg.discrete()}. 
In Section~\ref{sec:num_prob} we explore the different methods implemented in \code{bd.probability()} and \code{bdg.probability()}. 
Finally, Section~\ref{sec:num_est} utilizes the alternative parameter estimation frameworks and their underlying techniques. 
The code for these examples can be found at \href{https://github.com/bpatch/birdepy\_examples}{https://github.com/bpatch/birdepy\_examples}. 

\subsection{Simulation}\label{sec:num_sim}
\begin{figure}[h]
\centering
\begin{subfigure}[b]{.44\linewidth}
\centering
 \includegraphics{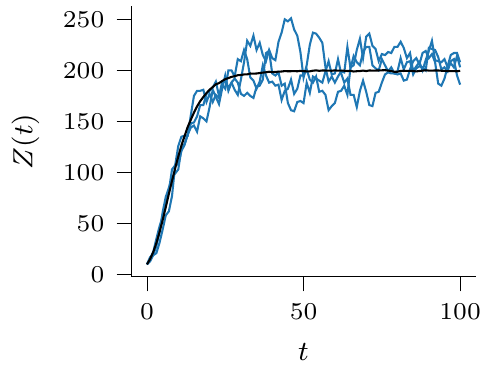}
\caption{Method set to \code{"exact"}.}
 \end{subfigure}
 \begin{subfigure}[b]{.44\linewidth}
 \centering
  \includegraphics{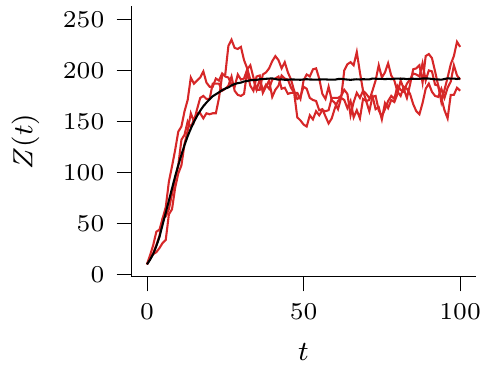}
\caption{Method set to \code{"ea"}.}
 \end{subfigure}

\begin{subfigure}[b]{.44\linewidth}
\centering
\vspace{5mm}
 \includegraphics{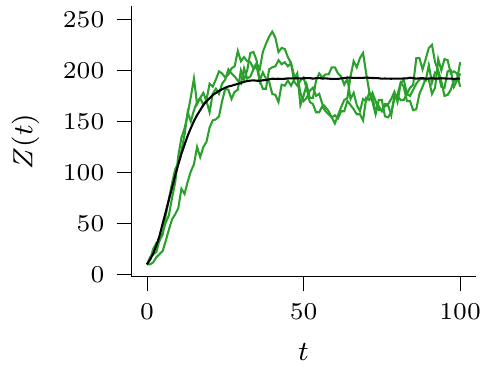}
\caption{Method set to \code{"ma"}.}
 \end{subfigure}
 \begin{subfigure}[b]{.44\linewidth}
 \centering
 \vspace{5mm}
  \includegraphics{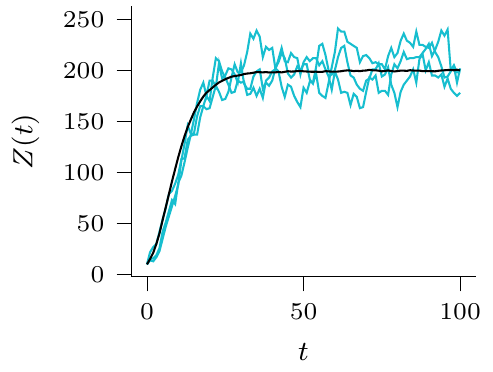}
\caption{Method set to \code{"gwa"}.}
 \end{subfigure}
\caption{Sample paths and mean approximations of a Hassell model with $\gamma=0.75$, $\nu=0.25$, $\alpha=0.01$ and $c=1$ generated using \code{bd.simulate.discrete()} with method set to: (a) \code{"exact"}, (b) \code{"ea"}, (c) \code{"ma"} and (d) \code{"gwa"}. }
\label{fig:Hassell_Trajectories}
\end{figure}

To illustrate the simulation capabilities of \pkg{BirDePy}, we consider a Hassell model with $\gamma=0.75$, $\nu=0.25$, $\alpha=0.01$  and $c=1$ (recall that a Hassell model with $c=1$ is also known as a Beverton--Holt model). 
Initializing with $Z(0)=10$, the four plots in Figure~\ref{fig:Hassell_Trajectories} each contain three simulated sample paths of the process, along with a mean curve estimated from $10^3$ sample paths. 
The difference between each plot is the simulation algorithm used to generate the sample paths. 
In all cases,
\begin{CodeInput}
bd.simulate.discrete([0.75, 0.25, 100, 1], "Hassell", 10, np.arange(0, 101, 1),
                     method=m, k=10**3, seed=2021) 
\end{CodeInput}
is executed, and for each plot (from left to right, and top to bottom), the value of \code{m} is set to \code{"exact"}, \code{"ea"}, \code{"ma"} and \code{"gwa"} (as described in Table~\ref{tab:sims}). 
Observe that due to differences in the way that random numbers are utilized by each simulation method, the sample paths are not identical, even though the same seed is used. 
Despite this, the three approximation methods \code{"ea"}, \code{"ma"} and \code{"gwa"} generate sample paths that appear highly similar to the one generated by the \code{"exact"} method. 

\begin{figure}[h]
\centering
\includegraphics[width=10cm]{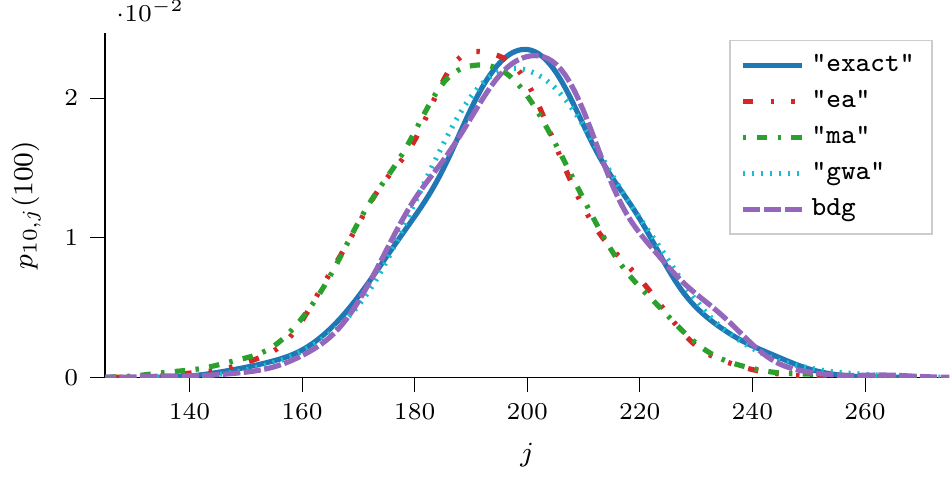}
\caption{Kernel density estimates for the distribution of $Z(100)$ conditional on $Z(0)=10$ for a Hassell model with $\gamma=0.75$, $\nu=0.25$, $\alpha=0.01$ and $c=1$, generated using different methods. }
\label{fig:sim_kde}
\end{figure}

Figure~\ref{fig:sim_kde} provides a more detailed examination of the simulated output at time $t=100$. 
The figure displays \textit{kernel density estimates} (KDEs) for the distribution of $Z(100)$ generated from $10^3$ samples. 
In addition to including output of the four algorithms implemented in \code{bd.simulate.discrete()}, the figure also includes a KDE generated using the output of executing 
\begin{CodeInput}
bdg.discrete([0.75, 0.25, 0.01, 1], "Hassell", 10, 100, k=10**3, seed=2021)
\end{CodeInput}
which provides samples generated using exact simulation performed on a GPU. 
This figure makes it clear that the approximated simulation methods can indeed differ notably from exact simulation. 
Despite the seed being the same and both using the same `exact' simulation algorithm, \code{bd.simulate.discrete(method="exact")} and \code{bdg.discrete()} generate different KDEs (for this sample size) due to the different way that each function handles random numbers. 
As displayed in Table~\ref{tab:sim_times}, the methods and functions also differ substantially in their computation time.  
So that these times can be fairly compared, \code{np.arange(0, 101, 1)} is replaced by \code{[0, 100]} in the execution of 
\code{bd.simulate.discrete()} for the purpose of generating Figure~\ref{fig:sim_kde}. 

\begin{table}[h]
\centering
\begin{tabular}{ccccc}
\hline
 \code{"exact"} &     \code{"ea"} &     \code{"ma"} &     \code{"gwa"} &    \code{bdg.discrete()} \\
\hline
 73 & 3.81 & 7.19 & 14 & 0.1090 \\
\hline
\end{tabular}
\caption{Computation time (CPU + GPU) in seconds used to generate the sample paths underlying the KDEs in Figure~\ref{fig:sim_kde}. }
\label{tab:sim_times}
\end{table}

\begin{figure}[h]
\centering
\includegraphics[width=\textwidth]{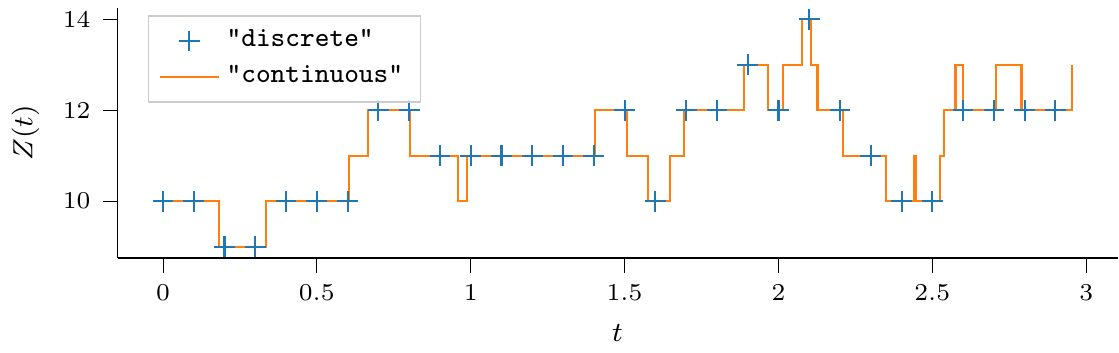}
\caption{Discretely-observed and continuously-observed simulated trajectories of a linear birth-and-death process with $\gamma=0.5$ and $\nu=0.45$ initiated with $Z(0)=10$.}
\label{fig:cts_sim}
\end{figure}

We conclude this section by comparing the output of 
\begin{CodeInput}
bd.simulate.discrete([0.5, 0.45], "linear", 10, np.arange(0, 3, 0.1),
                     seed=2021)
\end{CodeInput}
which generates a discretely-observed trajectory using exact simulation, and 
\begin{CodeInput}
bd.simulate.continuous([0.5, 0.45], "linear", 10, 3, seed=2021)
\end{CodeInput}
which generates a continuously-observed trajectory (also exactly). 
In both cases a linear birth-and-death process with $\gamma=0.5$ and $\nu=0.45$ initiated with $Z(0)=10$ is simulated over 3 units of time. 
The trajectories are plotted in Figure~\ref{fig:cts_sim}. 
Observe that in this case, setting the seeds equal ensures that the output from the two functions matches at the discrete observation times. 

\subsection{Transition probabilities}\label{sec:num_prob}
\begin{figure}[h!]
\centering
\begin{subfigure}[b]{.44\linewidth}
\centering
 \includegraphics{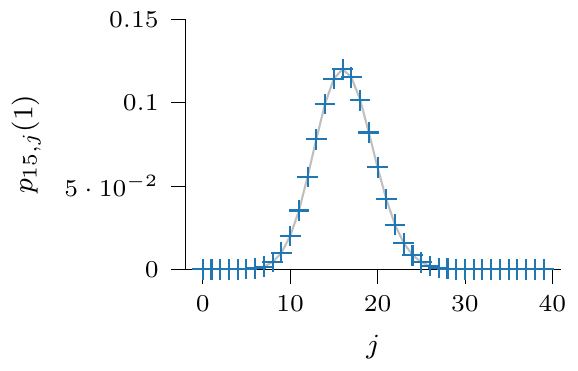}
\caption{\code{"expm"}.}
 \end{subfigure}
 \begin{subfigure}[b]{.44\linewidth}
 \centering
  \includegraphics{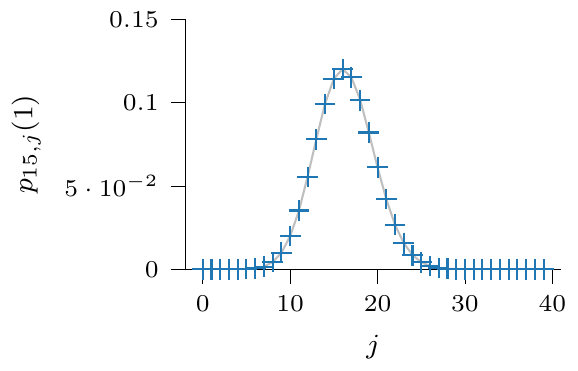}
\caption{\code{"uniform"}.}
 \end{subfigure}

 \begin{subfigure}[b]{.44\linewidth}
 \centering
  \includegraphics{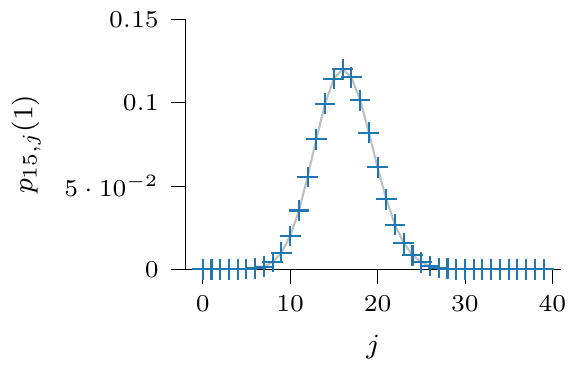}
\caption{\code{"Erlang"}.}
 \end{subfigure} 
 \begin{subfigure}[b]{.44\linewidth}
 \centering
  \includegraphics{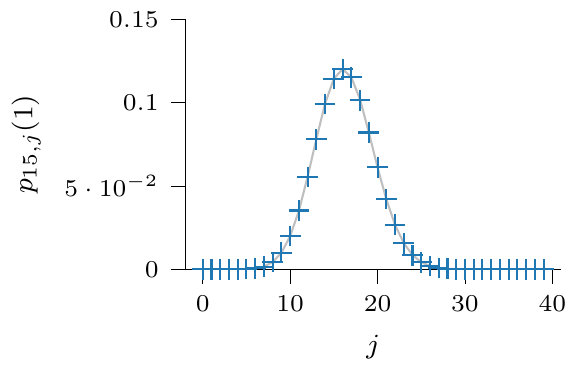}
\caption{\code{"ilt"}.}
 \end{subfigure} 

 \begin{subfigure}[b]{.44\linewidth}
 \centering
  \includegraphics{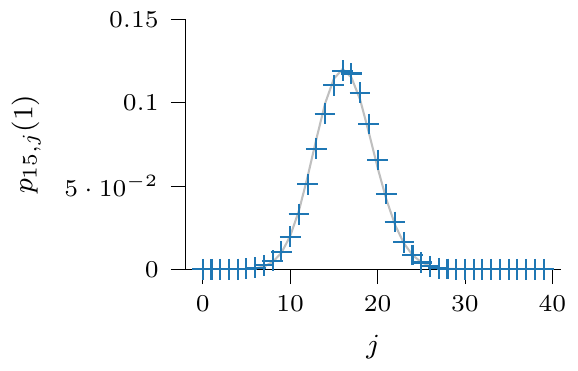}
\caption{\code{"da"}.}
 \end{subfigure}
  \begin{subfigure}[b]{.44\linewidth}
 \centering
  \includegraphics{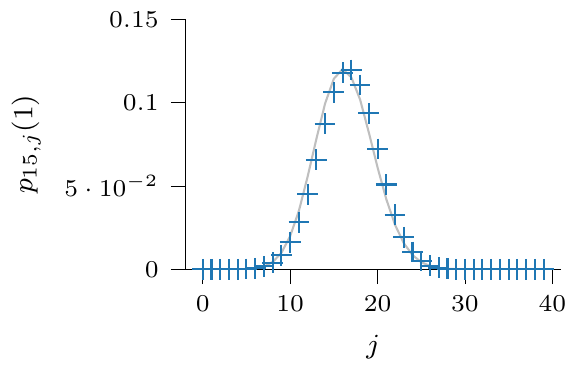}
\caption{\code{"oua"}.}
 \end{subfigure}

  \begin{subfigure}[b]{.44\linewidth}
 \centering
  \includegraphics{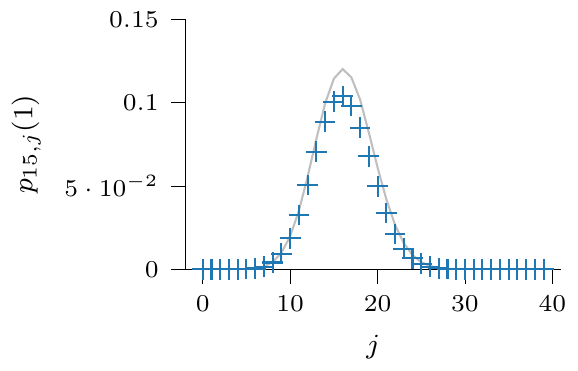}
\caption{\code{"gwa"}.}
 \end{subfigure}
   \begin{subfigure}[b]{.44\linewidth}
 \centering
  \includegraphics{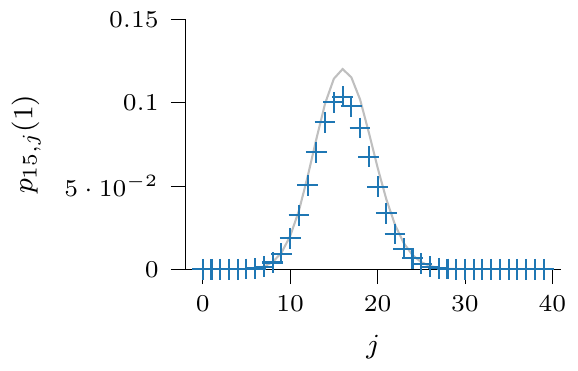}
\caption{\code{"gwasa"}.}
 \end{subfigure}
\caption{Transition probabilities $p_{15,j}(1)=\mathbb P(Z(1)=j\mid Z(0)=15)$ for the Verhulst model with $\gamma=0.8$, $\nu=0.4$, $\alpha=0.025$ and $\beta=0$, approximated using simulation (gray) and as outputted by \code{bd.probability()} (blue crosses) with method set to: (a) \code{"expm"}, (b) \code{"uniform"}, (c) \code{"Erlang"}, (d) \code{"ilt"}, (e) \code{"da"}, (f) \code{"oua"}, (g) \code{"gwa"}, and (h) \code{"gwasa"}. }
\label{fig:prob_small_verhulst}
\end{figure}

In this section, we illustrate how to compute approximate transition probabilities $p_{i,j}(t)$ using \pkg{BirDePy}. 
Consider a Verhulst model with $\gamma=0.8$, $\nu=0.4$, $\alpha=0.025$, and $\beta=0$. 
This model is also studied in \cite{ross2006parameter}. 
Figure~\ref{fig:prob_small_verhulst} displays approximations to $p_{15,j}(1)=\mathbb P(Z(1)=j\mid Z(0)=15)$ for  $0\leq j < 40$, as outputted by executing
\begin{CodeInput}
bd.probability(15, np.arange(0, 40, 1), 1, [0.8, 0.4, 0.025, 0], 
               model="Verhulst", method=m)
\end{CodeInput}
with \code{m} taking on the values \code{"expm"}, \code{"uniform"}, \code{"Erlang"}, \code{"ilt"}, \code{"da"}, \code{"oua"}, \code{"gwa"} and \code{"gwasa"} (as described in Table~\ref{tab:probs}). 
In the background of each plot is a density generated by executing
\begin{CodeInput}
bdg.probability(15, np.arange(0, 40, 1), 1, [0.8, 0.4, 0.025, 0], 
                model="Verhulst", k=10**6)
\end{CodeInput}
which provides a simulation-based approximation for verification of the reliability of the output. 
All of the methods provide accurate transition probability approximations. 
The methods \code{"gwa"} and \code{"gwasa"} exhibit slight inaccuracy around the mode of the distribution. 
Table~\ref{tab:small_verhulst_times} reports the computation times needed to generate each of the plots in Figure~\ref{fig:prob_small_verhulst}. 
These times vary greatly, with \code{"gwasa"} being the fastest and \code{"ilt"} being the slowest.

\begin{table}[h!]
\centering
\begin{tabular}{ccccccccc}
\hline
\code{bdg}  &    \code{"expm"}  &    \code{"uniform"}  &    \code{"Erlang"}  &      \code{"ilt"}  &      \code{"da"}  &     \code{"oua"}  &     \code{"gwa"}  &    \code{"gwasa"}  \\
\hline
 0.5840 & 0.0070 &    0.7700 &   0.0030 & 2.05 & 0.0050 & 0.0040 & 0.0250 &  0.0030 \\
\hline
\end{tabular}
\caption{Computation time (CPU + GPU) in seconds used to generate the densities in Figure~\ref{fig:prob_small_verhulst}. }
\label{tab:small_verhulst_times}
\end{table}

\subsection{Estimation}\label{sec:num_est}
We next demonstrate how to use the wide variety of estimation methods implemented in \pkg{BirDePy}. 
Consider a Verhulst model with $\gamma=0.8$, $\nu=0.4$, $\alpha=0.025$, and $\beta=0$. 
As per Section~\ref{sec:num_sim}, executing
\begin{CodeInput}
obs_times = list(range(100))
param = [0.8, 0.4, 0.025, 0]
p_data = bd.simulate.discrete(param, 'Verhulst', 5, obs_times, seed=2021, k=5)
t_data = [obs_times for _ in range(num_sample_paths)]
\end{CodeInput}
generates synthetic data for this model, which is used to generate all estimates in this section. 
This synthetic data consists of five sample paths, each with 100 equally-spaced observations. 

We first use the \code{"abc"}  framework of \code{bd.estimate()}, which utilizes the ABC algorithm, to estimate $\gamma$ from \code{t_data} and \code{p_data}. 
The usage of this framework is described in Section~\ref{sec:bd.estimate_abc}. 
Executing 
\begin{CodeInput}
est = bd.estimate(t_data, p_data, [0.5], [[0,1]], framework='abc',
                  model='Verhulst', idx_known_p=[1, 2, 3], max_its=1,
                  known_p=[0.4, 0.025, 0], seed=2021)
\end{CodeInput}
results in an estimate of $\gamma$ of 0.7468 being stored in the variable \code{est.p}, and a standard error of 0.1034 being stored in the variable \code{est.se}. 
To obtain this estimate we have specified 0.5 as an initial guess for the value of $\gamma$ and placed bounds $[0, 1]$ on the estimate. 
The estimate took 74 seconds to be generated.
Note that the above code utilises the option \code{max_its=1}, meaning that this estimate is obtained after a single basic ABC iteration is performed. 
Removing this option (so that the default \code{max_its=3} is utilised instead) results in a $\gamma$ estimate equal to 0.7620 with standard error of 0.0533. 
By using more iterations the estimate becomes slightly more accurate and the standard error becomes substantially smaller. 
This improvement in accuracy brings with it a higher computational burden --- in this case the estimate took 1853 seconds to be generated. 
This time is very large relative to other experiments we conducted using this framework with different models, however it indicates the framework is not working efficiently for all PSDBDPs. For this example we found that this framework failed to compute reasonable estimates of $\gamma$, unless values of $\nu$, $\alpha$ and $\beta$ were assumed to be known. 
These factors suggest that more research needs to be performed on developing ABC algorithms tailored to PSDBDPs. 

Following this, we use the \code{"dnm"}  framework of \code{bd.estimate()} which directly (numerically) maximizes approximations to the likelihood function. 
As described in Section~\ref{sec:bd.estimate_dnm}, when using this framework, several  methods can be used to generate an approximation to the likelihood function, as determined by the argument of the \code{likelihood} parameter. 
For this framework, and in the following discussion of frameworks \code{"em"} and \code{"lse"}, we prepare to use \code{bd.estimate()} by first executing
\begin{CodeInput}
con = {'type': 'ineq', 'fun': lambda p: p[0]-p[1]}
alpha_max = 1/np.amax(np.array(p_data))
alpha_mid = 0.5*alpha_max
\end{CodeInput}
which sets the constraint $\gamma>\nu$, provides an upper bound for $\alpha$ based on the maximum observed population, and uses this bound to provide an initial guess for $\alpha$. 
We also set \code{opt_method} to \code{"differential-evolution"} as this produces more reliable estimates (at the cost of taking longer to compute them). 
Table~\ref{tab:dnm_estimates} displays the estimates obtained for each parameter, and the corresponding asymptotic standard errors obtained by executing 
\begin{CodeInput}
est = bd.estimate(t_data, p_data, [0.51, 0.5, alpha_mid], 
                  [[1e-6,5], [1e-6,5], [1e-6, alpha_max]], model='Verhulst',
                  framework='dnm', known_p=[0], idx_known_p=[3], 
                  con=con, likelihood=m, opt_method='differential-evolution')
\end{CodeInput}
for different choices of likelihood approximation method \code{m}.  
Many of the methods return accurate estimates. 
The final column of Table~\ref{tab:dnm_estimates} reports the CPU times needed to obtain the estimates, which exhibit substantial variability. 
The Erlang likelihood method appears to be both the fastest and among the most accurate of the implemented methods for this example when using the \code{"dnm"} framework. 
In our experience, Erlang performs well in many cases, and at times when it does not perform well, increasing the argument of parameter \code{k} reliably improves the performance (here \code{k=150}, which is its default value). 

\begin{table}[h!]
\centering
\begin{tabular}{lcccc}
\hline
    Likelihood method (\code{m})    &   $\gamma$ &     $\nu$ &   $\alpha$  & Time (secs)\\
\hline
 \code{"da"}      &  0.7817 (0.0612) & 0.3906 (0.0292) &  0.0252 (0.0013) & 90\\
 \code{"Erlang"}  &  0.7824 (0.0653) & 0.3873 (0.0294) &  0.0250 (0.0013) & 0.8610\\
 \code{"expm"}    &  0.7810 (0.0650) & 0.3867 (0.0292) &  0.0250 (0.0013) & 1.1\\
 \code{"gwa"}     &  0.4111 (0.0432) & 0.3414 (0.0228) &  0.0076 (0.0041) & 88\\
 \code{"gwasa"}   &  0.4198 (0.0438) & 0.3494 (0.0239) &  0.0075 (0.0040) & 7.39\\
 \code{"ilt"}     &  0.7782 (0.0047) & 0.3852 (0.0052) &  0.0251 ~~~(*)~~~~   & 4738\\
 \code{"oua"}     &  0.6528 (0.0570) & 0.3716 (0.0298) &  0.0229 (0.0016) & 16\\
 \code{"uniform"} &  0.7811 (0.0650) & 0.3867 (0.0292) &  0.0250 (0.0013) & 2.85 \\
\hline
\end{tabular}
\caption{Estimates of $(\gamma, \nu, \alpha)$ and standard errors (in brackets), along with the corresponding CPU times, for a Verhulst model with $\gamma=0.8$, $\nu=0.4$, and $\alpha=0.025$ (with $\beta=0$ assumed to be known), generated using \code{bd.estimate(framework="dnm")} with different methods for approximating the likelihood function.}
\label{tab:dnm_estimates}
\end{table}

Moving on, we use the \code{"em"} framework of \code{bd.estimate()} which implements EM algorithm approaches to obtaining maximum likelihood estimates. 
This framework has multiple choices for parameters \code{technique} and \code{accelerator} available in \pkg{BirDePy}, as detailed in Section~\ref{sec:bd.estimate_em}. 
The estimates, standard errors, and computational times are displayed in Table~\ref{tab:em}, and again exhibit a great deal of variability. 
The EM algorithm accelerated by the method of \cite{lange1995quasi} with the integrals in Equation~\eqref{eq:em_integrals} evaluated using matrix exponentials appears to be highly accurate and efficiently computed. 
The corresponding estimate $(0.8031, 0.4039, 0.0246)$ appears to outperform any estimate computed using the \code{"dnm"} framework at only a small cost in computation time of 4.44 seconds. 

\begin{table}[h!]
\centering
\begin{tabular}{llcccc}
\hline
  Technique    &   Accelerator    &   $\gamma$ &     $\nu$ &   $\alpha$  & Time (secs) \\
\hline
 \code{"expm"} & \code{"cg"}    &  0.7796 (0.0649) & 0.3865 (0.0292) &  0.0250 (0.0013) &                4.66\\
 \code{"expm"} & \code{"none"}  &  0.8175 (0.0712) & 0.4079 (0.0325) &  0.0248 (0.0013) &                4.36\\
 \code{"expm"} & \code{"Lange"} &  0.8031 (0.0700) & 0.4039 (0.0318) &  0.0246 (0.0013) &                4.44\\
 \code{"expm"} & \code{"qn1"}   &  0.7810 (0.0650) & 0.3867 (0.0292) &  0.0250 (0.0013) &               11.02\\
 \code{"expm"} & \code{"qn2"}   &  0.7526 (0.0655) & 0.4068 (0.0322) &  0.0231 (0.0015) &                3.6\\
 \code{"ilt"}  & \code{"cg"}    &  0.7815 (0.0649) & 0.3862 (0.0289) &  0.0250 (0.0012) &             9410\\
 \code{"ilt"}  & \code{"none"}  &  0.8171 (0.0139) & 0.4078 (0.0138) &  0.0248 (0.0012) &             8339\\
 \code{"ilt"}  & \code{"Lange"} &  0.8060 (0.0378) & 0.4025 (0.0305) &  0.0248 ~~~(*)~~~~&             98595199\\
 \code{"ilt"}  & \code{"qn1"}   &  0.7802 (0.0649) & 0.3864 (0.0290) &  0.0250 (0.0013) &            19995\\
 \code{"ilt"}  & \code{"qn2"}   &  0.7535 (0.0592) & 0.4073 (0.0326) &  0.0238 (0.0013) &             6198\\
 \code{"num"}  & \code{"cg"}    &  0.8626 (0.0778) & 0.4296 (0.0363) &  0.0249 (0.0012) &               12\\
 \code{"num"}  & \code{"none"}  &  0.6153 (0.0490) & 0.4414 (0.0423) &  0.0172 (0.0018) &               88\\
 \code{"num"}  & \code{"Lange"} &  0.7859 (0.0656) & 0.3895 (0.0296) &  0.0250 (0.0013) &               27\\
 \code{"num"}  & \code{"qn1"}   &  0.5046 ~~~(*)~~~~& 0.5039 ~~~(*)~~~~&  0.0161 ~~~(*)~~~~&               47\\
 \code{"num"}  & \code{"qn2"}   &  0.6609 (0.0625) & 0.4275 (0.0352) &  0.0184 (0.0020) &               9.83\\
\hline
\end{tabular}
\caption{Estimates $(\gamma, \nu, \alpha)$ and standard errors (in brackets), along with the corresponding CPU time, for a Verhulst model with $\gamma=0.8$, $\nu=0.4$, and $\alpha=0.025$ (with $\beta=0$ assumed to be known), generated using \code{bd.estimate(framework="em")} with different techniques for evaluating the integrals in Equation~\eqref{eq:em_integrals} and different accelerators.}
\label{tab:em}
\end{table}

Finally, we use the \code{"lse"} framework of \code{bd.estimate()}. 
The estimates, standard errors, and computational times are displayed in Table~\ref{tab:lse}, and are again highly variable, although not to the same scale as the \code{"dnm"} and \code{"em"} values. 
If a user is prepared to go without standard errors, the computation time is only a fraction of the displayed values, as the CPU times shown in the table are mostly consumed determining standard errors. 
For our example, this framework has not produced results which are as accurate as those produced by the other frameworks. 

\begin{table}[h!]
\centering
\begin{tabular}{lcccc}
\hline
   Squares       &   $\gamma$ &     $\nu$ &   $\alpha$ &   Time (secs)\\
\hline
  \code{"expm"} &  0.7078 (0.0383) & 0.2988 (0.0241) &  0.0288 (0.0021)  &               21 \\
  \code{"fm"}   &  0.7109 (0.0367) & 0.2955 (0.0257) &  0.0279 (0.0025)  &              485 \\
  \code{"gwa"}  &  0.6792 (0.0480) & 0.3278 (0.0354) &  0.0244 (0.0024)  &               16 \\
\hline
\end{tabular}
\caption{Estimates of $(\gamma, \nu, \alpha)$ and standard errors (in brackets), along with the corresponding CPU time, for a Verhulst model with $\gamma=0.8$, $\nu=0.4$, and $\alpha=0.025$ (with $\beta=0$ assumed to be known), generated using \code{bd.estimate(framework="lse")} with different methods for evaluating the squared error.}
\label{tab:lse}
\end{table}

\section{Case studies}\label{sec:case}
In this section we consider two discretely-observed endangered bird populations. 
Section~\ref{sec:robins} examines the black robin population on Rangatira Island in New Zealand, and Section~\ref{sec:cranes} studies whooping cranes which migrate annually between Canada's Wood Buffalo National Park (WBNP) and the Aransas National Wildlife Refuge (ANWR) in Texas. 
In general, \pkg{BirDePy} is intended as a decision support tool to be used in conjunction with expert opinion and combined with other analyses. 
These case studies are intended to illustrate the role that our package could play as part of a broader analysis of these bird populations. 
The code used for these case studies can be found at \href{https://github.com/bpatch/birdepy\_examples}{https://github.com/bpatch/birdepy\_examples}. 

\subsection{Chatham Island black robins}\label{sec:robins} 
\begin{figure}[h]
\centering
\begin{subfigure}[b]{.49\linewidth}
\centering
 \includegraphics[width=6cm]{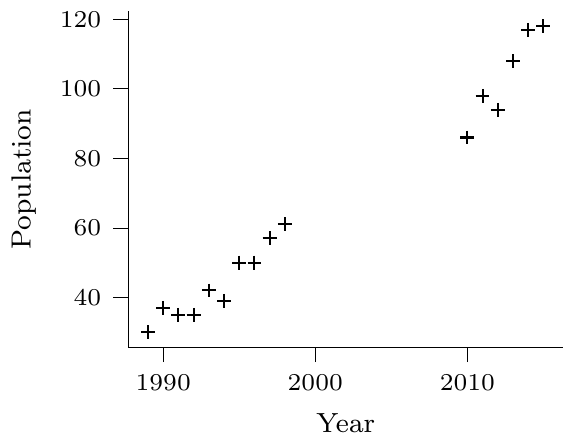}
\caption{}
\label{fig:pop_counts_robin}
 \end{subfigure}
 \begin{subfigure}[b]{.49\linewidth}
 \centering
  \includegraphics[width=6cm]{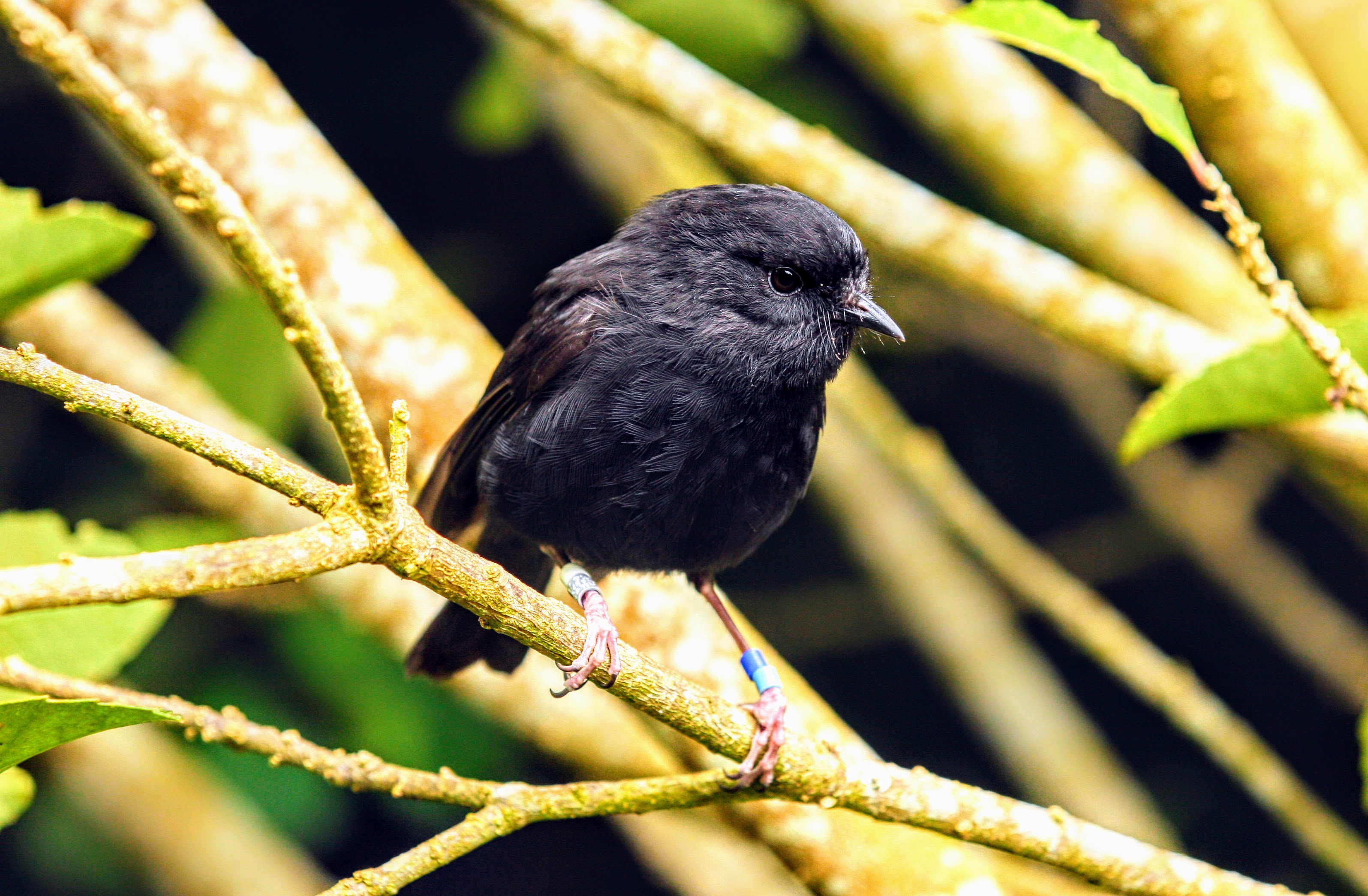}
  \vspace{7mm}
\caption{}
\label{fig:robin}
 \end{subfigure}
 \caption{(a) Yearly population counts for the female black robins on Rangatira Island, and (b) A black robin on Rangatira Island (photo by {\em Melanie Massaro}).}
\end{figure}

As reported by \cite{ButlerMerton1992}, the Chatham Island black robin ({\em Petroica traversi}) population started to come under threat approximately 450-500 years ago when humans arrived at the Chatham Islands. 
This was primarily due to the introduction of cats and rats. 
By 1893, the species had been completely wiped out on the main island of the Chatham Islands, with the remaining 20-35 surviving birds found on Little Mangere Island. 
This population persisted for the following nine decades (1893-1976), but experienced a sharp fall in the period 1972-1976. 
The only surviving seven birds, which included just one breeding female, were relocated to a safer habitat on Mangere Island in 1976. 
Out of the five birds remaining in 1979, only one breeding pair produced offspring that bred successfully \cite{kennedy2009extinction}. 
A period of intensive management followed until the spring of 1990. 
As part of this management, a second population of black robins was established on Rangatira Island. 
This isolated population has not been subject to management, but has still been observed sporadically over the last few decades. 
Figure~\ref{fig:pop_counts_robin} displays annual counts of the females in this population for 1989-1998 and 2010-2015 \citep{massaro2013nest,davison2021parameter}. 
The gap in observation between 1999 and 2009 is due to a lack of government funding for this period \citep{massaro2013nest}. 
Note that \code{bd.estimate()} does not require discretely-observed population counts to be evenly spaced, so this gap in observation is handled well by the function. 
Executing
\begin{CodeInput}
t_data = [1989, 1990, 1991, 1992, 1993, 1994, 1995, 1996, 1997, 1998, 2010,
          2011, 2012, 2013, 2014, 2015]
p_data = [30, 37, 35, 35, 42, 39, 50, 50, 57, 61, 86, 98, 94, 108, 117, 118]
\end{CodeInput}
is a simple way to input this data into \proglang{Python}, although other methods where the data is first stored in a spreadsheet or other database are also possible. 

Deep-forest interiors and dense forest-edges are known to provide suitable habitats for black robins to successfully breed in.
Furthermore, since black robins are territorial, each breeding pair typically occupies a habitat patch that cannot be shared with other breeding pairs.  
When the availability of deep-forest interiors and dense forest-edges becomes limited, black robins are not particularly suited to expanding their territory into shrubland or scattered vegetation \cite{kennedy2009extinction}. 
This type of limitation on the capacity of a population to expand is a pervasive occurrence in ecology (e.g., \cite{brook2000predictive,ford2002selection,hilderbrand2003roles}). 

The lack of suitable habitat frequently slows the growth of populations of endangered species. Resources are restricted in remaining habitat patches, and therefore a population can grow only until it reaches the maximum population size a particular habitat can support, which is a biological definition of the \textit{carrying capacity}. The knowledge of the carrying capacity can help policy makers to set realistic goals for population expansion within a particular habitat, and to devise conservation strategies.

Using \pkg{BirDePy} we are able to fit the data displayed in Figure~\ref{fig:pop_counts_robin} to the models in Table~\ref{tab:rates} and, consequently, to find estimates of the carrying capacity of Rangatira Island. 
For PSDBDPs, the carrying capacity corresponds to the closest integer to $z^\star>0$ such that $\lambda_{z^\star}=\mu_{z^\star}$. 
As an example, for a Ricker model $z^\star = \frac{1}{\alpha}\big(\log(\gamma/\nu)\big)^{1/c}$. 
The output of \code{bd.estimate()} stores an estimate of the carrying capacity in the attribute \code{capacity}.  

In our analysis, we consider two Verhulst models, a Ricker model, a B-H model, a Hassell model, an MS-S model, and the linear model. 
In particular, our first Verhulst model attributes any decline in population growth to restrictions on the birth rate (by assuming $\beta=0$), while our second Verhulst model assumes that an increase in mortality is the cause (by assuming $\alpha=0$). 
Given the limited available data, we found that \code{bd.estimate()} performed better when only three parameters needed to be estimated. 
We therefore set $c=1$ for the Ricker model, and $c=2$ for the MS-S and Hassell models. 
We do not consider a linear-migration model for this population since it is known that black robins do not migrate between the Chatham Islands. 

To perform estimation for Verhulst 1, we execute
\begin{CodeInput}
est = bd.estimate(t_data, p_data, [2, 2, 0.05], [[0,10], [0,10], [0, 1]],
                  model="Verhulst",  known_p=[0], idx_known_p=[3],
                  opt_method="differential-evolution", seed=2021)
\end{CodeInput}
and then parameter estimates, standard errors, the carrying capacity estimate, and the likelihood are stored, respectively, in \code{est.p}, \code{est.se}, \code{est.capacity}, and \code{np.exp(est.val)}. 
Observe that we initialize with a parameter guess of \code{[2, 2, 0.05]} and instate parameter bounds \code{[[0, 10], [0, 10], [0,1]]}. 
We also set \code{opt_method} to \code{"differential-evolution"}, as this produces more reliable estimates (at the cost of a long computation time). 
We also execute similar code to obtain results for the other models.

The results are presented in Table~\ref{tab:est_robin}. 
The B-H model has the highest likelihood (1.02e-21), and suggests that the population (of females) is subject to a carrying capacity of 146. 
The other PSDBDPs are not very different from the B-H model in terms of likelihood (9.27e-22 to 1.01e-21), and suggest a similar carrying capacity of 133 to 143. 
These figures are comparable to the upper limit on the carrying capacity of 170 breeding pairs that is reported by \cite{massaro2018post}. 
It is worth highlighting that while our estimates are based solely on population size counts, the estimate in \cite{massaro2018post} requires  precise knowledge of the habitat that is costly to acquire. 

The black robin data we are using was previously fitted to a linear birth-and-death process in \cite{davison2021parameter}. We see that the linear model has substantially lower likelihood (5.59e-22) than the PSDBDP models, which provides further evidence of the existence of a carrying capacity. 
Despite this, it should be noted that none of the PSDBDP models have $\alpha$ or $\beta$ which is more than two standard errors from zero. 
This means that, despite the strong indication that a carrying capacity exists given by the  higher likelihood of the PSDBDP models, there is not enough evidence in the data to reject a null hypothesis of $\alpha=\beta=0$. 

\begin{figure}[h]
\begin{subfigure}[b]{.45\linewidth}
\includegraphics[width=8cm]{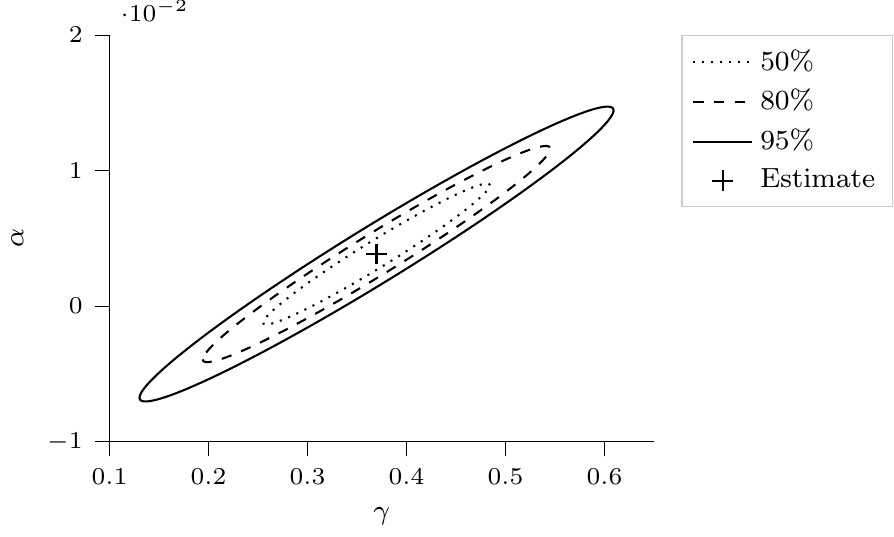}
\caption{Asymptotic confidence region.}
 \end{subfigure}
 \hspace{10mm}
 \begin{subfigure}[b]{.45\linewidth}
 \includegraphics[width=8cm]{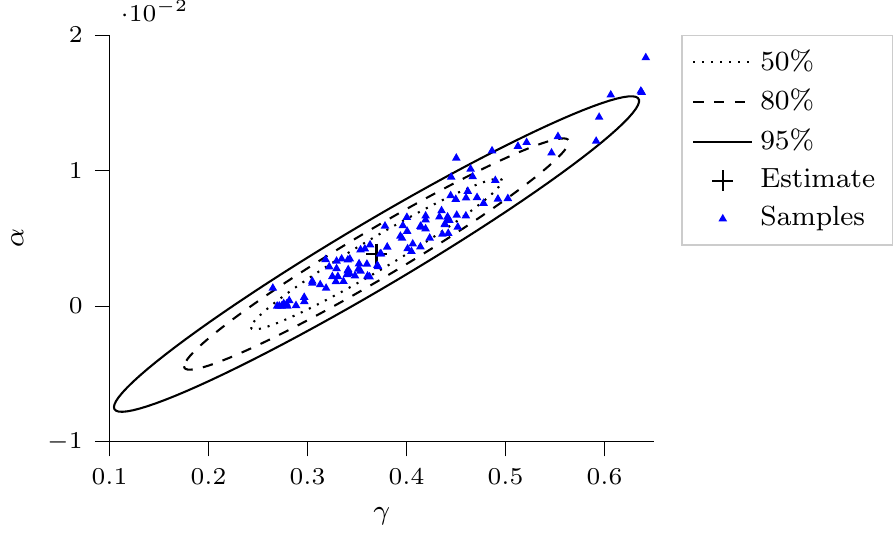}
 \caption{Simulated confidence region.}
 \end{subfigure}
\caption{Asympototic (left) and simulated (right) confidence regions for $(\gamma, \alpha)$ assuming $\nu=0.2367$ for a B-H model fitted to the black robin population data displayed in Figure~\ref{fig:pop_counts_robin}.}
\label{fig:ci_robin}
\end{figure}

Obtaining confidence regions for the model parameters requires us to fix the value of one of the three parameters, since \pkg{BirDePy} cannot plot confidence regions consisting of more than two parameters. Here we assume that $\nu=0.2367$, which is the B-H model estimate for this parameter (comparable to the death rate estimate in \cite{hautphenne2019fitting}). Then, executing
\begin{CodeInput}
bd.estimate(t_data_b, p_data_b, [0.36, 0.0017], [[0,1], [0, 0.1]],
            model="Hassell", known_p=[0.2367, 1], idx_known_p=[1, 3], 
            se_type=s, ci_plot=True, seed=2021, xlabel='$\gamma$', 
            ylabel='$\\alpha$')
\end{CodeInput}
with \code{s="asymptotic"}, and then again with \code{s="simulated"}, results in the confidence regions for $(\gamma, \alpha)$ displayed in Figure~\ref{fig:ci_robin}, which show the likely range of values that $(\gamma, \alpha)$ take  conditional on $\nu=0.2367$. 
Note that here we have set an initial condition close to the estimated parameter value and returned to the default optimization routine to speed up the computation of the simulated confidence region. 

\begin{figure}[h]
\centering
 \includegraphics{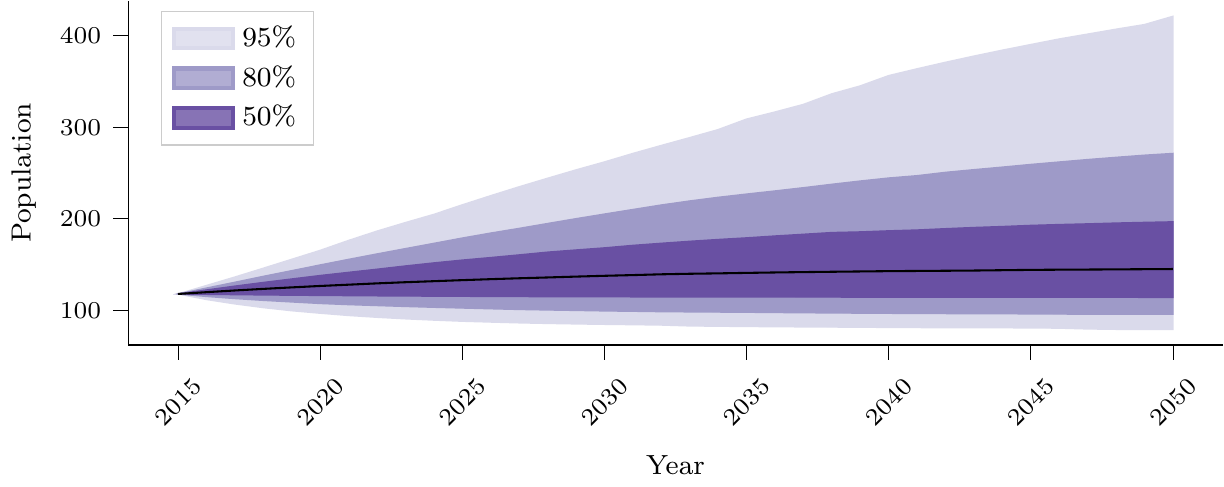}
  \caption{Confidence intervals for the expected future population of female black robins on Rangatira Island as generated by \code{bd.forecast()} assuming a B-H model.}
\label{fig:forecast_robin}
\end{figure}

Figure~\ref{fig:forecast_robin} shows the effect of uncertainty in the value of $(\gamma, \nu, \alpha)$ on the likely range of values taken by the expected future population size. 
We see that large increases in population size may occur, but the bulk of the projected population sizes are close to the carrying capacity estimates discussed earlier. 
Executing 
\begin{CodeInput}
bd.forecast("Hassell", p_data[-1], np.arange(2015, 2051, 1), est.p, 
            cov=est.cov,  p_bounds=[[0,10], [0,10], [0, 1]], 
            known_p=[1], idx_known_p=[3])
\end{CodeInput}
produces this figure. 

\begin{table}[h!]
\centering
\begin{tabular}{lccccc}
\hline
 Model      &   $\gamma$ &     $\nu$ &   $\alpha$ or $\beta$  & Capacity           &   Likelihood \\
\hline
 Verhulst 1    &  0.3505 (0.1299) & 0.2382 (0.1000) &       0.0023 (0.0018) & 138 &  9.84e-22 \\
 Verhulst 2    &  0.3018 (0.1051) & 0.1860 (0.1036) &       0.0046 (0.0059) & 135 & 9.57e-22 \\
 Ricker        &  0.3587 (0.1365) & 0.2380 (0.1002) &       0.0029 (0.0027) & 142 &  9.99e-22 \\
 B-H           &  0.3690 (0.1475) & 0.2367 (0.0998) &       0.0038 (0.0045) & 146 &  1.02e-21\\
 Hassell       &  0.3687 (0.1446) & 0.2418 (0.1034) &       0.0016 (0.0017) & 143 &  1.01e-21\\
 MS-S          &  0.3283 (0.1190) & 0.2413 (0.1019) &       0.0045 (0.0026) & 133 &  9.27e-22\\
 linear        &  0.2845 (0.0957) & 0.2350 (0.0956) &        -      & - & 5.59e-22\\
\hline
\end{tabular}
\caption{Parameter estimates and standard errors (in brackets), along with their corresponding carrying capacity values and likelihoods, as determined by \code{bd.estimate()} using the black robin data displayed in Figure~\ref{fig:pop_counts_robin}. }
\label{tab:est_robin}
\end{table}

Finally we note that many PSDBDPs, including models with a carrying capacity, eventually become extinct with probability one (typically after a very long time), that is, $\mathbb P(\lim_{t\to\infty}Z(t)=0)=1$. Fitting a single population trajectory to these models (such as in the case of the black robins) then induces a small bias in the parameters estimates. We refer to \cite{braunsteins2020parameter,braunsteins2021parameter} for discussions and results on consistency and asymptotic normality of estimators for (discrete-time) Markov population processes with almost sure extinction. 

\subsection{Whooping cranes}\label{sec:cranes}
According to \cite{Allen1952cranes}, there were approximately 1300-1400 whooping cranes ({\em Grus americana}) in existence around 1860-1870. 
In the first part of the 20th century, hunting and a loss of natural habitat to agriculture pushed whooping cranes very close to extinction, with only 36 birds in 1912, declining down to 15 birds in 1941 \cite{Allen1952cranes}. 
The Migratory Bird Act of 1916, and the establishment of the WBNP in 1922 as the primary summer habitat and the ANWR in 1937 as the primary winter habitat, are thought to be major contributors to the survival of the species \cite{national2005endangered}. 
In 1967, whooping cranes were one of 75 species placed on the inaugural US endangered species list and put under federal protection. 
The Whooping Crane Recovery Plan 2006 \cite{CranePlan2006} states that one way the species can be down-listed from `endangered' to `threatened' is for the ANWR flock to reach a self-sustaining population above 1000 individuals.

\begin{figure}[h]
\centering
\begin{subfigure}[b]{.49\linewidth}
\centering
 \includegraphics{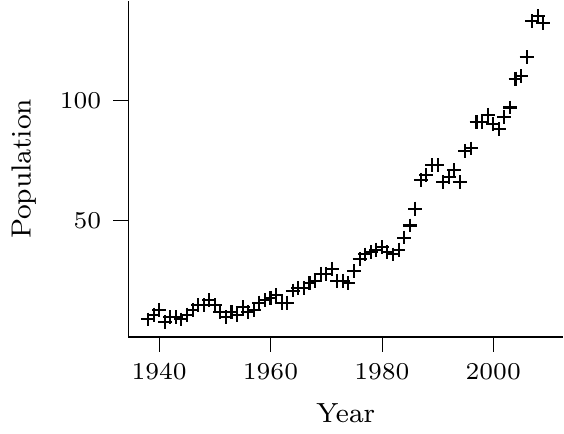}
\caption{}
\label{fig:pop_counts_crane}
 \end{subfigure}
 \begin{subfigure}[b]{.49\linewidth}
 \centering
 \includegraphics[width=6cm]{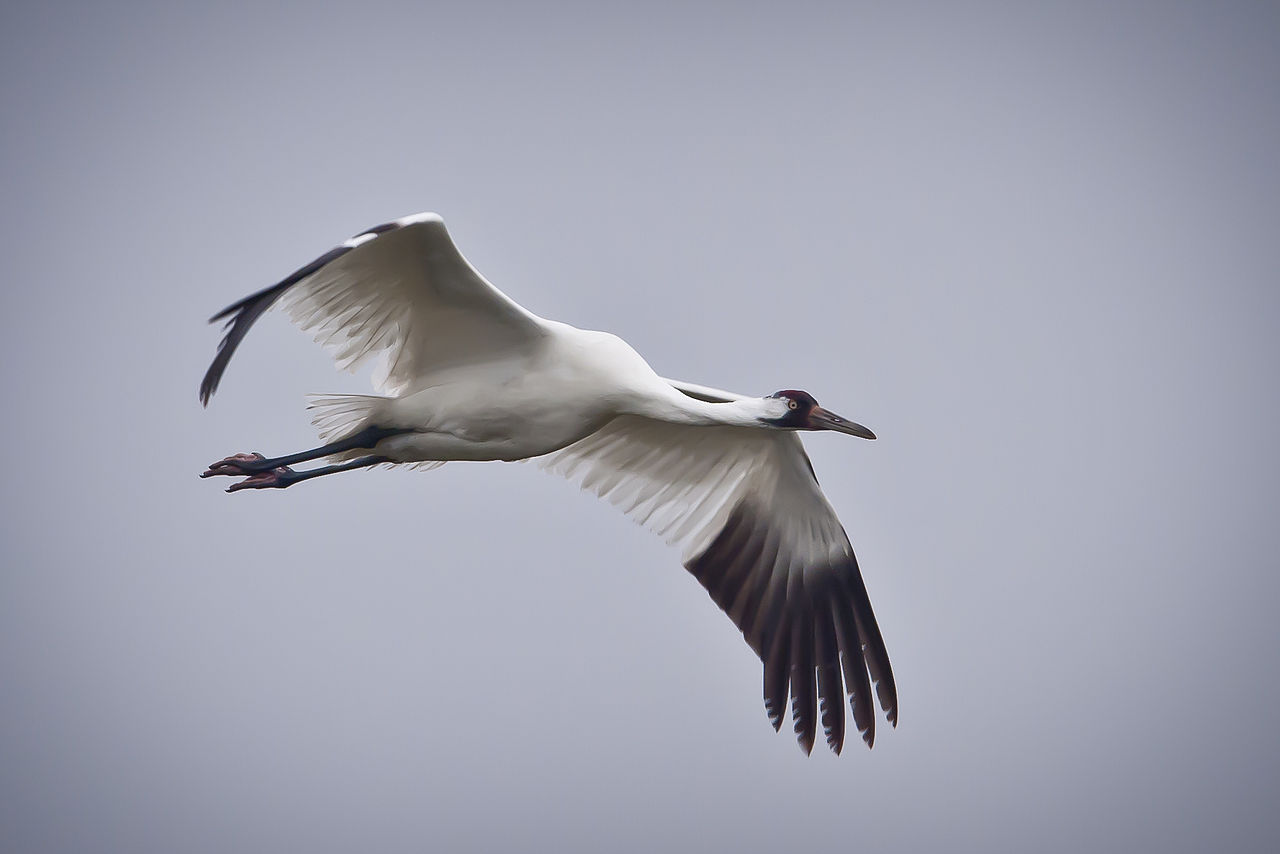}
 \vspace{10mm}
\caption{}
\label{fig:crane}
 \end{subfigure}
 \caption{(a) Assumed yearly population counts for female whooping cranes, and (b) ``Whooping Crane in flight in Texas'' by {\em U.S.\ Department of Agriculture} is licensed under \href{https://creativecommons.org/licenses/by/2.0/}{CC BY 2.0}.}
\end{figure}

Estimates of the annual population in the ANWR flock for 1938-2009 are provided in \cite{butler2013influence}. 
Following \cite{davison2021parameter} (where a subset of this data is also analyzed), in Figure~\ref{fig:pop_counts_crane} we display estimates of the female population only, assuming a sex ratio of 1:1. 
Note that focusing on the female population eliminates dependencies between males and females due to reproduction, and makes birth-and-death models more suitable. Executing
\begin{CodeInput}
t_data = [t for t in range(1938, 2010, 1)]
p_data = [9, 11, 13, 8, 10, 10, 9, 11, 13, 15, 15, 17, 15, 12, 10, 12, 11,
          14, 12, 13, 16, 17, 18, 19, 16, 16, 21, 22, 22, 24, 25, 28, 28, 
          30, 25, 25, 24, 29, 34, 36, 37, 38, 39, 37, 36, 38, 43, 48, 55,  
          67, 69, 73, 73, 66, 68, 71, 66, 79, 80, 91, 91, 94, 90, 88, 93, 
          97, 109, 110, 118, 133, 135, 132]
\end{CodeInput}
inputs this data into \proglang{Python}. 
A large portion of the data was obtained using aerial surveys, as detailed by \cite{strobel2014monitoring}. 

Our analysis considers the same models as those described in Section~\ref{sec:robins} with the addition of a linear-migration model. 
Similar to Section~\ref{sec:robins}, the \pkg{BirDePy} function \code{bd.estimate()} can be used to obtain parameter estimates, standard errors, carrying capacity estimates, and likelihoods. 
The results are shown in Table~\ref{tab:est_crane}. 
The linear-migration model has the highest likelihood (1.63e-81), suggesting that the population is not subject to a carrying capacity and is growing exponentially. 
Interestingly, however, the PSDBDP models have likelihoods in the range 1.50e-81 to 1.56e-81, which are: (i) relatively close to the linear-migration likelihood, and (ii) greater than the linear model likelihood (1.30e-81), which is the lowest among the considered models. 
This means that if our assumption of possible external migration into the population does not hold (meaning $\alpha=0$), then we would conversely conclude that the population is in fact subject to a carrying capacity. 
Therefore our assumption that external migration into the population is possible is crucial to determining whether a carrying capacity exists. 
In contrast to the black robin population, it is far less clear in this case whether the population is subject to a carrying capacity. 
The PSDBDP models suggest a carrying capacity of 223 females (likelihood 1.56e-81) to 411 females (likelihood 1.50e-81). 
This is in line with the analysis of \cite{stehn2010changes}, who estimate the current habitat can support up to 576 individual whooping cranes, with the possibility of up to 1156 if a nearby habitat is suitable for colonization and successfully colonized. 
Therefore our analysis suggests that there is cause for doubt about the ability of the population to achieve a threatened status by reaching a self-sustaining population of 1000 individuals in the ANWR flock. 

\begin{figure}[h]
\begin{subfigure}[b]{.4\linewidth{}}
\includegraphics[width=8cm]{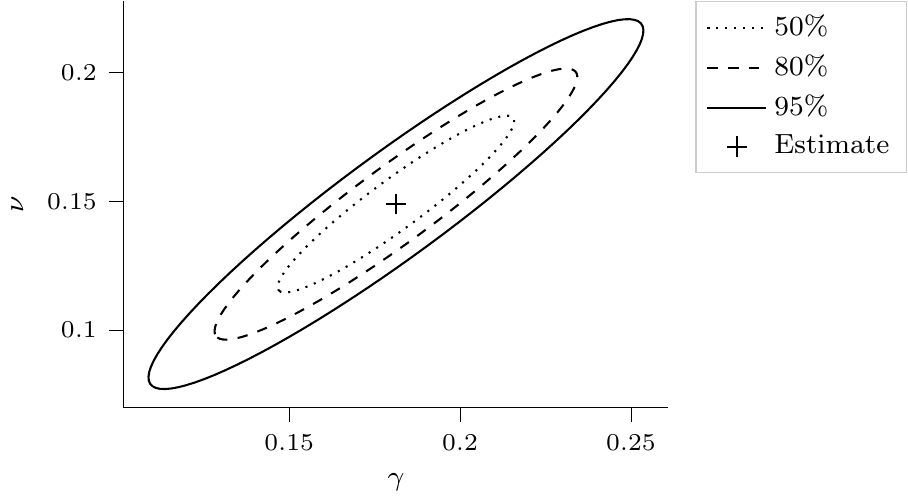}
 \caption{Asymptotic confidence region.}
 \end{subfigure}
 \hspace{10mm}
 \begin{subfigure}[b]{.4\linewidth}
 \hspace{5mm}
\includegraphics[width=8cm]{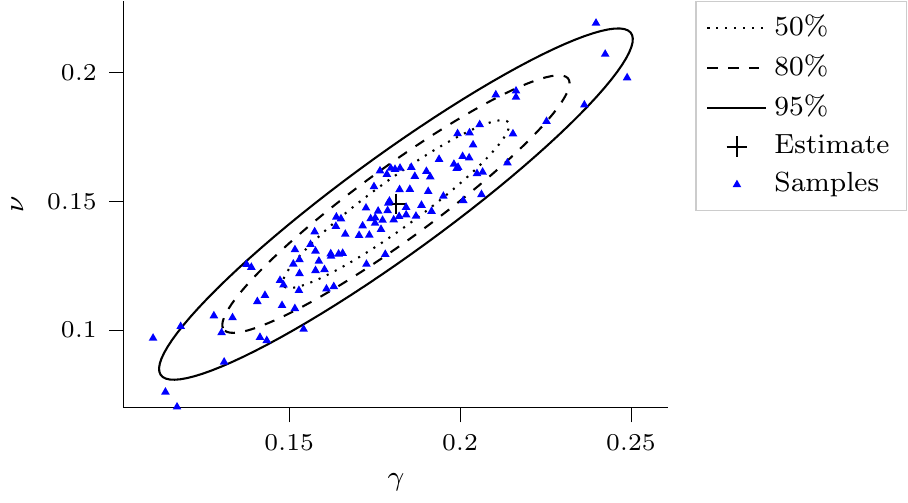}
 \caption{Simulated confidence region.}
 \end{subfigure}
\caption{Asymptotic (left) and simulated (right) confidence regions for $(\gamma, \nu)$ assuming $\alpha=0.3157$ for a linear-migration model fitted to the whopping crane data displayed in Figure~\ref{fig:pop_counts_crane}.}
\label{fig:ci_crane}
\end{figure}

\begin{figure}[h]
\centering
 \begin{subfigure}[b]{.99\linewidth}
 \centering
\includegraphics{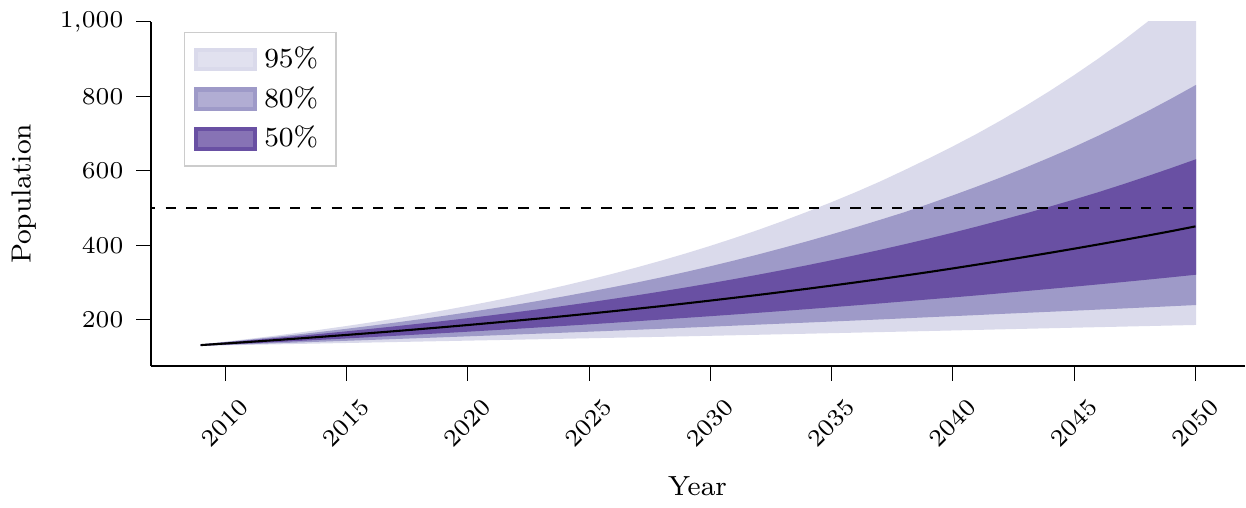}
 \caption{Linear-migration model.}
\end{subfigure}

 \begin{subfigure}[b]{.99\linewidth}
 \centering
\includegraphics{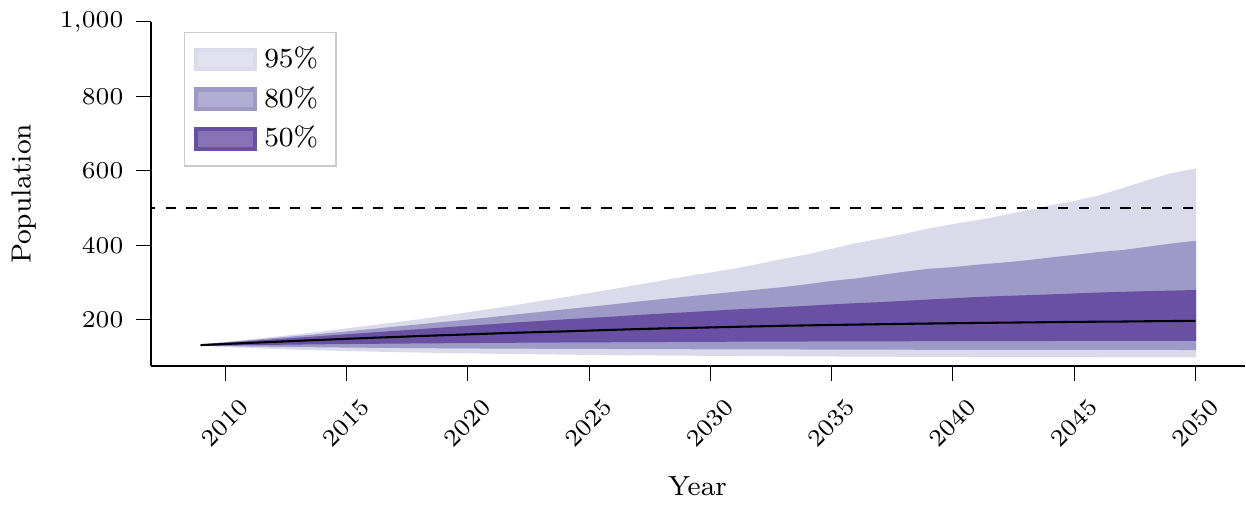}
 \caption{MS-S model.}
\end{subfigure}
\caption{Confidence intervals for the mean future population of female whopping cranes as generated by \code{bd.forecast()} assuming (a) a linear-migration model, and (b) an MS-S model. The dashed line shows the target population }
\label{fig:forecast_crane}
\end{figure}

The Verhulst 1 model (likelihood 1.51e-81) and the Verhulst 2 model (likelihood 1.52e-81) have highly similar likelihoods, so our analysis does not provide evidence in either direction as to whether a decline in the birth rate or an increase in the death rate is limiting population growth (assuming such a limit exists). 
Similar to Section~\ref{sec:robins}, as displayed in Figure~\ref{fig:ci_crane}, we are able to generate asymptotic and simulated confidence regions for $(\gamma, \nu)$ in the linear-migration model assuming $\alpha=0.3157$. 
Figure~\ref{fig:forecast_crane} displays how this uncertainty and the choice of model affects the likely range of values that the expected future population size takes. 
Assuming a linear-migration model, more than 50\% of values within the confidence interval exceed 500 by the year 2050, suggesting a long term outlook that may see whooping cranes removed from the endangered species list. 
On the other hand, if a MS-S model is more appropriate, then the portion of values within the confidence interval exceeding 500 by the year 2050 is substantially less, suggesting whooping cranes will continue to be endangered for some time to come under present conditions. 
This highlights the fact that the choice of model is fundamental in order to draw biological conclusions, which cannot be made without the additional input of experts. 

\begin{table}[h!]
\centering
\begin{tabular}{lccccc}
\hline
 Model            &   $\gamma$ &     $\nu$ &   $\alpha$ or $\beta$ & Capacity           &   Likelihood \\
\hline
 Verhulst 1       &  0.1998 (0.0351) & 0.1492 (0.0293) &       0.0008 (0.0013) & 330  &  1.51e-81 \\
 Verhulst 2       &  0.1931 (0.0303) & 0.1423 (0.0321) &       0.0011 (0.0021) & 325  &  1.52e-81  \\
 Ricker           &  0.1999 (0.0354) & 0.1493 (0.0293) &       0.0008 (0.0015) & 367  &  1.51e-81  \\
 Beverton-Holt    &  0.1999 (0.0357) & 0.1493 (0.0293) &       0.0008 (0.0016) & 411  &  1.50e-81 \\
 Hassell          &  0.1999 (0.0356) & 0.1493 (0.0293) &       0.0004 (0.0008) & 388  &  1.50e-81 \\
 MS-S             &  0.1966 (0.0320) & 0.1493 (0.0293) &       0.0025 (0.0022) & 223  &  1.56e-81 \\
 linear           &  0.1902 (0.0295) & 0.1506 (0.0293) &          -            & -    &  1.30e-81 \\
 linear-migration &  0.1812 (0.0317) & 0.1489 (0.0294) &       0.3157 (0.4768) & -    &  1.63e-81 \\
\hline
\end{tabular}
\caption{Parameter estimates and standard errors (in brackets), along with their corresponding carrying capacity values and likelihoods, as determined by \code{bd.estimate()} using the whooping crane population data displayed in Figure~\ref{fig:pop_counts_crane}. }
\label{tab:est_crane}
\end{table}

\section{Concluding remarks and future work} \label{sec:future}
\pkg{BirDePy} provides a collection of functions for working with PSDBDPs in \proglang{Python}. 
The ability of the package to approximate transition probabilities, to generate sample paths, to estimate parameters from discretely-observed population size data, and to forecast future population sizes has been illustrated with a variety of examples and through two pertinent case studies. 
We have demonstrated how this package can be of benefit to ecologists. 

\pkg{BirDePy} is available from the Python Package Index (PyPI) \href{https://pypi.org/project/birdepy/}{https://pypi.org/project/birdepy/} and is developed on GitHub \href{https://github.com/BirDePy}{https://github.com/BirDePy}, where contributions in the form of pull requests are welcomed. 

There are several models and features which would be natural for future versions of \pkg{BirDePy} to include:
\begin{itemize}
\item Currently, \pkg{BirDePy} is designed to work with PSDBDPs where individuals are homogeneous. 
However, many population processes involve individuals of multiple types. 
Key examples include epidemic models such as the Susceptible-Infectious-Recovered (SIR) model, queueing network models where individuals are classified by the server they are currently located at, and ecological models where individuals are classified according to their location or age. 
The package could be enhanced to accommodate these multi-type models. 
\item The transition rates of the PSDBDPs in \pkg{BirDePy} currently depend only on the population size. 
This could be expanded to a generalized framework that allows other covariates to influence birth and death rates. 
For example, if each transition could be associated with a vector of covariates, then the birth and death rates could be modeled as functions of regression coefficients in a generalized-linear-model framework.
The birth and death functions would be augmented to depend on the covariate vector, in addition to the parameters they currently depend upon. 
Such a framework would provide insights into which covariates are important determinants of birth and death rates. 
A framework along these lines is discussed by \cite{crawford2014estimation}. 
\item In our case studies we showed how the likelihood of the parametrized models and expert opinion could be combined to choose between the models in Table~\ref{tab:rates}. 
It would be useful if a more sophisticated method of model selection could be developed and incorporated into \pkg{BirDePy}. 
\item Each of the parameter estimation and transition probability approximation methods implemented in \pkg{BirDePy} have strengths and weaknesses depending on the model, parameter values and population size at hand. 
More research needs to be conducted on when each method is optimal to use. 
This would be particularly useful since then an ``ensemble'' method could be developed that switches automatically between the methods we have already included. 
\end{itemize}



\section*{Computational details}
The numerical experiments in this paper were all performed in \proglang{Python} (v3.7.10) with \pkg{BirDePy} (v0.0.9) on a desktop computer running Windows 10 with an Intel i5-10400F CPU, Nvidia GTX 1070 GPU, and 16GB RAM. 
\section*{Acknowledgments}
This research is funded by the Australian Government through the Australian Research Council (DP200101281).


 \bibliography{refs}


\newpage

\begin{appendix}

\end{appendix}


\end{document}